\documentclass[11pt]{article}
\usepackage[colorlinks=true,citecolor=blue,linkcolor=blue]{hyperref}
\usepackage{multirow}% for tables
\usepackage[left=2cm,right=2cm,top=2cm,bottom=2cm]{geometry}
\usepackage{cite}
\usepackage{psfrag}
\usepackage{graphicx}
\usepackage{epsfig}
\usepackage{booktabs}% for tables
\usepackage{amsmath,amsfonts,amssymb}
\usepackage{pstcol,pst-fill,pst-grad}
\usepackage{pstricks,pst-fill,pst-grad}
\usepackage{euscript}
\usepackage{pstricks}
\usepackage{wrapfig}
\usepackage{slashed}
\usepackage[toc,page]{appendix}
\usepackage{here}
\date{}
\begin{document}
\title{\vspace{-3cm}
		\hfill\parbox{4cm}{\normalsize \emph{}}\\
		\vspace{1cm}
		{Laser-assisted electron-nucleon scattering}}
	\vspace{2cm}

\author{I. Dahiri$^{1}$, M. Baouahi$^{1}$, M. Ouali$^1$, B. Manaut$^{1}$, E. Hrour$^{1,2}$, M. El Idrissi$^{1,3}$, S. Taj$^{1,}$\thanks{Corresponding author, E-mail: s.taj@usms.ma}  \\
		{\it {\small$^1$ Polydisciplinary Faculty, Laboratory of Research in Physics and Engineering Sciences,}}\\ {\it{\small Team of Modern and Applied Physics, Sultan Moulay Slimane University, Beni Mellal, 23000, Morocco.}} \\{\it {\small$^2$ Superior School of Technology of Beni Mellal, Sultan Moulay Slimane University, Beni Mellal, Morocco.}}\\ {\it {\small$^3$
Polydisciplinary Faculty of Khouribga, Sultan Moulay Slimane University, Beni Mellal, Morocco.}}
		}
\maketitle \setcounter{page}{1}
\date{\today}

\begin{abstract}
In both absence and presence of a circularly polarized monochromatic electromagnetic pulse, we have analyzed the electron-nucleon scattering process, where the nucleon is assumed to be spinless with a spherical shape. We have provided the theoretical calculation of the differential cross section (DCS) by using the Dirac-Volkov formalism. This research paper aims to provide two comparisons: We first compare the DCS in the absence of the laser field with its corresponding laser-assisted DCS. A second comparison is made between the electron-proton and electron-neutron scattering processes to study the effect of the laser on both processes. The results obtained about the effect of the laser field on the DCS and the electric form factor have been discussed for both scattering processes. We have found that the DCS is reduced when the laser field is applied for both processes. In addition, the form factor is also decreased by raising the incident electron energy in electron-proton scattering, but it increases in electron-neutron scattering. Moreover, the form factors for both scattering situations are unchanged by raising the laser field strength up to $10^{8}\,V/cm$. 
\end{abstract}
\maketitle
\section{Introduction}
The theory of quantum electrodynamics (QED) is one of the most successful physical theories, which describes how electrically charged particles interact by exchanging photons. It provides the simplest way to open up the domain of photon-matter interaction that is still unexplored completely and will ultimately enable the creation of new physics that makes use of extremely brilliant high-energy sources \cite{E144_Collaboration,Burke,Bamber,Mourou}. Extreme light physics is a new field that has benefited greatly from the experimental development of ultrashort lasers by LUXE at DESY \cite{Altarelli} and E320 at FACET-II (SLAC) \cite{FACET_II}. 
The latter, which were introduced in the 1960s \cite{Maiman}, has made it possible for researchers to comprehend the physical processes that are assisted or induced by the laser field. The creation of all the most powerful lasers was greatly aided by the chirped pulse amplification (CPA) \cite{Strickland} invented in 1985 by Donna Strickland and Gérard Mourou who awarded the Nobel Prize in Physics for this major scientific achievement. In addition to these experimental developments and breakthroughs, a lot of theoretical research has been done \cite{Attaourti1,Manaut2,taj1,taj2,idrissi}.
For example, there have been discussions concerning the possibility that the presence of a powerful laser pulse may have had a considerable impact on the nuclear $\beta$ decay \cite{Akhmedov,Reiss,Becker}.
Other theoretical investigations have suggested that a laser field with circular polarization decreases the cross section and the decay width of scattering \cite{Attaourti,Manaut3,Manaut4,Ouali1,Ouhammou1,ElAsri,Ouali2,Ouhammou2,Mekaoui} and decay \cite{Mouslih1,Jakha1,Jakha2,Mouslih2,Mouslih3,Baouahi1,Baouahi2} processes, respectively.
In 1996, Bula et al., made the first observation \cite{E144_Collaboration} of Compton multiphoton scattering \cite{Boca,Heinzl,Seipt,Mackenroth} in which one electron may absorb up to four photons from the laser field. 
In scattering processes in which an electron is elastically scattered by a target inside a laser field, the kinetic energy of the electron varies as a function of the photons' energy $\ell\hbar\omega$, where $\ell$ is an integer $(\ell=0, \pm1 ...)$ and $\omega$ is the frequency of the laser field. 
We refer to this scattering process, in the case of an atomic target, as LAES: Laser-Assisted Elastic Electron Scattering  (See the references \cite{Mittleman,Faisal,Mason,Ehlotzky,Joachain} for more details).
Using a cw-CO$_{2}$ laser beam, Andrick and Langhans reported the first observation of $\hbar\omega$ energy transfers in LAES in 1976 \cite{Andrick}. One year later, Weingartshofer et al. reported the multiphoton phenomena in electron-argon collisions \cite{Weingartshofer}, but this time by using a pulsed CO$_{2}$ laser field.

One of the physical processes envisaged in the presence of a laser field is the electron-proton scattering \cite{Dahiri1,Dahiri2} where the proton is considered as a structureless particle in its theoretical framework. 
In fact, electrons beam have been used to examine the structure of the proton and neutron, with the unique electron-proton collider HERA (Hadron Electron Ring Accelerator) at DESY, producing results that have been very crucial for understanding these nucleons. To solve the mystery of the proton charge radius, some experiments have been conducted elsewhere except at HERA. 
This charge radius can be calculated by measuring the cross section of the electron-proton elastic scattering or by using high-resolution spectroscopy of the hydrogen atom. The electron-proton cross section was measured in 2010 by the A1 collaboration at the Mainz Mikrotron (MAMI) to obtain new precise data about the proton charge radius \cite{Bernauer}. In the same year, the CREMA (Charge radius Experiment with Muonic Atom) collaboration realized an experiment using hydrogen spectroscopy of muons, and this allowed them to provide much more precise results \cite{Pohl}. Not only just the electric charge radius which is introduced during the electromagnetic interaction in the electron-proton elastic scattering, but the proton seems to be more democratic since it also has a weak charge, which violates the parity at JLab ( Thomas Jefferson National Accelerator Facility) \cite{QWEAK_Collaboration,Androic}. These investigations provide direct criteria about the nature of protons and neutrons composites, as well as the measurements of quark and gluon distributions inside the nucleon. In $e^{-} p\rightarrow e^{-} p$ scattering, the nature of the virtual photon-proton interaction depends strongly on the wavelength $\lambda=h/|\vec{q}|$ (where $|\vec{q}|$ represents the momentum transfer) for different energy scales of the electron. At very low electron energies $\lambda\gg r_{p}$ ($r_{p}$ is the radius of the proton), the proton is considered as a point-like object during the electron-proton elastic scattering, and this was the concept of an already performed work \cite{Dahiri1,Dahiri2}. In these papers, the scattering of a point-like proton by electron impact (polarized and unpolarized electrons) in the absence and presence of a laser field was studied. At low energies, $\lambda$ is proportional to the radius of the proton $r_{p}$, and this makes the scattered proton as an extended charged object. When we increase the energy of the incident electron, the wavelength will be short so that $\lambda< r_{p}$, and the proton will be seen as an object composed of quarks. In addition, at very high electron energies $\lambda\ll r_{p}$, the proton appears as a sea of quarks and gluons. Because the proton is made up of quarks in the two last cases, it is no longer considered as a Dirac particle.

Therefore, this paper aims to add the influence of the laser field during electron-nucleon scattering when $\lambda$ is comparable to the nucleon radius. In order to analyze the influence of the laser field on the electric form factor, the nucleon will be treated as a spinless particle during this scattering process. The laser field's impact will be examined only for charged electrons using the Volkov formalism \cite{Volkov} for unpolarized electrons. Plane-wave functions will be used to represent initial and final electrons. In the second section, we construct a theoretical analytical calculation of the differential cross section of electron-nucleon scattering in the laboratory reference system in both absence and presence of a laser field. We assume that, in the first Born approximation, the electron exchanges a virtual photon $\gamma$ or a $Z^{0}$ boson with the nucleon during the scattering process (Electroweak interaction). This theoretical approach can also be examined when the target is a system of protons and neutrons. In the third section, we present the different obtained results. Finally, a brief conclusion is given in the last section.
\section{Overview of the theory}\label{Sec.1}
The general scattering process of a spinless nucleon by an electron impact in the absence and presence of a laser field is examined in this paper. The following Feynman diagram describes this scattering process via photon or $Z^{0}$-boson exchange:
\begin{figure}[H]
\centering
    \includegraphics[scale=0.25]{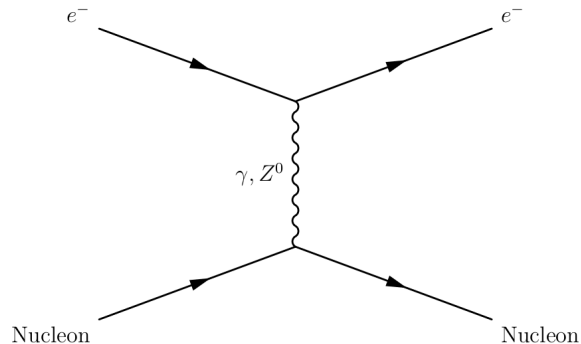}
   \caption{Feynman diagram for the electroweak electron-nucleon scattering ($\gamma$ photon and $Z^{0}$ boson exchange).} \label{Figure:0}
\end{figure}
We will start by giving the theoretical treatment of the cross section, which is the measurable parameter that defines the scattering process. In fact, when the incident electron acquires important energy, the nucleon is no longer considered as an elementary charge. In this case, the electric form factor is taken into consideration in the scattering process calculations.
In the first part, we will calculate the cross-section of this scattering process in the absence of a laser field. 
The corresponding form factor's formalism approach will be discussed in the next section. 
Then, we calculate the cross section of the studied process in the presence of a laser field.
Note that this work is developed for a space-time metric of signature $(+, -, -, -)$. In the theoretical calculation, we use the system of natural units $\hbar=c=1$.
\subsection{Laser-free differential cross section}
In the table below, we present the various parameters that describe both initial and final particles involved in the scattering process of a nucleon by electronic impact in the absence of the laser field:
\begin{table}[H]
\begin{center}
\begin{tabular}{|c||c|c|c|c|}
\toprule
 & Four-momentum & Energy & Momentum vector & Mass \\ \hline \toprule
Incident electron &~ $p_{1}$ ~~&~ $E_{1}$ &~~ $\vec{p_{1}}$ &~~ $m_{e}$  \\  \hline
Scattered electron &~ $p_{2}$ ~~&~ $E_{2}$ &~~ $\vec{p_{2}}$ &~~ $m_{e}$  \\ \hline
Initial nucleon &~ $p$  ~~&~ $E_{p}$ &~~ $\vec{p}$ &~~ $m_{N}$  \\ \hline
Final nucleon &~ $p'$ ~~&~ $E_{p'}$ &~~ $\vec{p'}$ &~~ $m_{N}$  \\ \hline
Four-momentum transfer ~~&~ $q=p-p'$ &~~ $E_{p}-E_{p'}$ &~ $\vec{q}=\vec{p}-\vec{p'}$ &~~ $-$   \\ \hline
 \toprule
\end{tabular}
\end{center}
\caption{Different laser-free parameters that describe both initial and final particles in electron-nucleon scattering.}\label{table1}
\end{table}
The scattering amplitude of electron-nucleon scattering by the exchange of a virtual photon, $\gamma$, is:
\begin{equation}\label{eq1}
 T_{\gamma}=e^{2}\bar{u}_{e}(p_{2},s_{2})\gamma^{\alpha}u_{e}(p_{1},s_{1})\dfrac{-g_{\alpha\beta}}{q^{2}}\left\langle p'| J^{\beta}_{em}(0)| p\right\rangle ,
\end{equation}
where $e$ is the electric charge of the electron. $u_{e}$ is a free Dirac bispinor for the electron, and it verifies the following normalization relation $\sum_{s_{i}}u_{e}(p_{i},s_{i})\bar{u}_{e}(p_{i},s_{i})=\slashed{p}_{i}+m_{e}$. 
The $J^{\beta}_{em}(0)$ operator corresponds to the elastic electromagnetic transition of the nucleon-nucleon current given by:
\begin{equation}\label{eq2}
\left\langle p'| J^{\beta}_{em}(0)| p\right\rangle=(p+p')^{\beta}ZF(q^{2}).
\end{equation}
In equation (\ref{Figure:1}), $Z$ is the proton number, and $F(q^{2})$ is the form factor normalized to $F(q^{2}=0)=1$. 
In the case of $Z^{0}$-boson exchange, the scattering amplitude can be expressed as follows:
\begin{eqnarray}\label{eq3}
T_{Z}&=&-\dfrac{e^{2}}{16\sin^{2}(\theta_{w})\cos^{2}(\theta_{w})}\bar{u}_{e}(p_{2},s_{2})[-\gamma^{\alpha}(1-4\sin^{2}(\theta_{w}))+\gamma^{\alpha}\gamma_{5}]u_{e}(p_{1},s_{1})\nonumber \\
&\times&\dfrac{-g_{\alpha\beta}+q_{\alpha}q_{\beta}/m^{2}_{Z^{0}}}{q^{2}-m^{2}_{Z^{0}}}\left\langle p'| K^{\beta}(0)| p\right\rangle,
\end{eqnarray}
where $\theta_{w}$ is the Weinberg angle, and $m_{Z^{0}}$ is the mass of the $Z^{0}$ boson. The $K^{\beta}(0)$ operator describes the elastic transition of the nucleon-nucleon current by the exchange of a $Z^{0}$ boson, and it is defined by:
\begin{eqnarray}\label{eq4}
\left\langle p'| K^{\beta}(0)| p\right\rangle=(p+p')^{\beta}[Z(1-4\sin^{2}(\theta_{w}))-N]F(q^{2}),
\end{eqnarray}
with $N$ denotes the number of neutrons in the nucleus. The terms $q_{\alpha}q_{\beta}$ of the $Z^{0}$-propagator in equation (\ref{eq3}) do not contribute. The sum of $T_{\gamma}$ and $T_{Z}$ can be expressed as follows:
\begin{eqnarray}\label{eq5}
 T_{\gamma}+T_{Z}&=&-e^{2}\dfrac{Z}{q^{2}}(p+p')_{\alpha}F(q^{2})\bar{u}_{e}(p_{2},s_{2})\gamma^{\alpha}\bigg[1+\dfrac{(1-4\sin^{2}(\theta_{w}))[(1-4\sin^{2}(\theta_{w}))-N/Z]}{16\sin^{2}(\theta_{w})\cos^{2}(\theta_{w})}\nonumber \\
 &\times&\dfrac{q^{2}}{q^{2}-m^{2}_{Z^{0}}}-\dfrac{[(1-4\sin^{2}(\theta_{w}))-N/Z]}{16\sin^{2}(\theta_{w})\cos^{2}(\theta_{w})}\dfrac{q^{2}}{q^{2}-m^{2}_{Z^{0}}}\gamma_{5}\bigg]u_{e}(p_{1},s_{1}).
\end{eqnarray}
For simplicity reasons, we use the following two abbreviations: $\textit{v}(q^{2})$ and $a(q^{2})$, such that:
\begin{eqnarray}\label{eq6}
\textit{v}(q^{2})&=&1+\dfrac{(1-4\sin^{2}(\theta_{w}))[(1-4\sin^{2}(\theta_{w}))-N/Z]}{16\sin^{2}(\theta_{w})\cos^{2}(\theta_{w})}\dfrac{q^{2}}{q^{2}-m^{2}_{Z}},\\ 
a(q^{2})&=&-\dfrac{[(1-4\sin^{2}(\theta_{w}))-N/Z]}{16\sin^{2}(\theta_{w})\cos^{2}(\theta_{w})}\dfrac{q^{2}}{q^{2}-m^{2}_{Z}}/\textit{v}(q^{2}).
\end{eqnarray}
In addition, we know that:
\begin{center}
$\begin{cases}
P_{+}=(1+\gamma_{5})/2 \\
P_{-}=(1-\gamma_{5})/2,
\end{cases}$
 $\qquad\text{with:}\qquad
\begin{cases}
P_{+}+P_{1}=1 \\
 P_{+}-P_{1}=\gamma_{5}.
 \end{cases}$
\end{center}
Therefore, $T_{\gamma}+T_{Z}$ is expressed in terms of $P_{+}$ and $P_{-}$ as follows:
\begin{eqnarray}\label{eq8}
T_{\gamma}+T_{Z}&=&-e^{2}\dfrac{Z}{q^{2}}(p+p')_{\alpha}F(q^{2})\textit{v}(q^{2})\bar{u}_{e}(p_{2},s_{2})\gamma^{\alpha}\bigg[(1+a(q^{2}))P_{+}+(1-a(q^{2}))P_{-}\bigg]u_{e}(p_{1},s_{1}).
\end{eqnarray}
The scattering matrix element $S_{fi}$ in the absence of the laser field is expressed by the following equation:
\begin{eqnarray}\label{eq9}
S_{fi}&=&\dfrac{(2\pi)^{4}\delta^{4}(p_{2}+p'-p_{1}-p)}{4V^{2}\sqrt{E_{1}E_{2}E_{p}E_{p'}}}( T_{\gamma}+T_{Z}),
\end{eqnarray}
where the Dirac-function $\delta^{4}(p_{2}+p'-p_{1}-p)$ ensures the conservation of the four-momentum through the scattering process. 
To obtain the cross-section we multiply the square of $S_{fi}$ by the phase space of the particles in the final state, and we divide it by the flux of the incident particles $|J_{inc}|=|v_{i}-V_{i}|/V$, by the observation time interval $T$, by the volume $V$, and by the number of target particles per unit volume $\rho=1/V$.
\begin{eqnarray}\label{eq10}
d\sigma&=&\dfrac{1}{VT}\dfrac{1}{|J_{inc}|}\dfrac{1}{\rho}\int\dfrac{Vd^{3}\vec{p_{2}}}{(2\pi)^{3}}\int\dfrac{Vd^{3}\vec{p'}}{(2\pi)^{3}}|S_{fi}|^{2}.
\end{eqnarray}
By following the mathematical analysis of $d\sigma$, the final expression of the differential cross section, denoted by the symbol $d\overline{\sigma}/d\Omega_{f}$, can be expressed as follows:
\begin{eqnarray}\label{eq11}
\frac{d\overline{\sigma}}{d\Omega_{f}}=\dfrac{Z^{2}e^{4}}{8(2\pi)^{2}m_{p}}\dfrac{|\vec{p_{2}}|}{|\vec{p_{1}}|}\dfrac{|F(q^{2})|^{2}|\textit{v}(q^{2})|^{2}}{|q|^{4}}\dfrac{|\overline{\mathcal{M}_{fi}}|^{2}}{g'(E_{2})},
\end{eqnarray}
where the $\overline{\mathcal{M}_{fi}}$ element represents the electron-nucleon scattering amplitude given by:
\begin{eqnarray}\label{eq12}
|\overline{\mathcal{M}_{fi}}|^{2}&=&\dfrac{1}{2}Tr[(\slashed{p}_{2}+m_{e})(\slashed{p}+\slashed{p}')[(1+a(q^{2}))P_{+}+(1-a(q^{2}))P_{-}](\slashed{p}_{1}+m_{e}) \nonumber \\
&\times &\bigg[(1+a(q^{2}))P_{-}+(1-a(q^{2}))P_{+}\bigg](\slashed{p}+\slashed{p}')].
\end{eqnarray}
In equation (\ref{eq12}), the factor $1/2$ is generated by averaging over the electron's initial polarization. Besides, the $ g'(E_{2}) $ function is the derivative of $g(E_{2})=(p+p_{1}-p_{2})^{2}-m_{N}^{2}$ with respect to $E_{2}$.
\subsection{The form factor}
For larger values of $|\vec{q}|$, the reduced wavelength of the virtual boson propagator decreases and the resolution increases. Consequently, the scattered electron no longer sees the total charge of the proton as in Mott scattering, but only a part of it. For this purpose, the spatial extension of the nucleon is described by a form factor $F(q^{2})$, which contains all the information about the spatial distribution of the charge. 
By using the Fourier transformation of the charge distribution function $\rho(x)$, the form factor $F(q^{2})$ can be expressed in the static approximation by the following expression: 
\begin{eqnarray}\label{eq13}
F(q^{2})&=&\int e^{iq.x}\rho(r)d^{3}r \nonumber \\
&\simeq&1-\dfrac{1}{6}\left\langle r^{2}\right\rangle q^{2}.
\end{eqnarray}
The following normalizing criterion of $\rho$ is used during the form factor development:
\begin{eqnarray}\label{eq14}
\int\rho(r)d^{3}r=4\pi\int_{0}^{+\infty}\rho(r)r^{2}dr=1.
\end{eqnarray} 
\subsection{Laser-assisted differential cross section}
Particles inside a circularly polarized laser field acquire an effective momentum. Therefore, we give in table \ref{table2} the new parameters that describe the final and initial particles. The proton is considered as a free particle and keeps the same parameters as in the previous section. 
The circular polarization of the electromagnetic field is defined by the four-vector potential $ A^{\mu}$ such that $ A^{\mu}(\phi)=a_{1}^{\mu}\cos(\phi)+a_{2}^{\mu}\sin(\phi) $. This potential verifies the Lorentz gauge (the condition of transversality $ k.A=0 $), with $ \phi=k.x $ is a phase and $k=(\omega,\vec{k})$ denotes the electromagnetic wave four-vector. 
The four polarizations $ a_{1}^{\mu}=|a|(0,1,0,0) $ and $ a_{2}^{\mu}=|a|(0,0,1,0) $ are orthogonal such that $ a_{1}^{2}=a_{2}^{2}=a^{2}=-\vert a\vert^{2}=-(\dfrac{\varepsilon_{0}}{\omega})^{2} $ where $\varepsilon_{0}$ is the laser field strength.
\begin{table}[H]
\begin{center}
\begin{tabular}{|c||c|c|c|c|}
\toprule
Electron & Four-momentum & Energy & Momentum vector & Mass \\ \hline \toprule
Incident  &~ $q_{1}=p_{1}-\big[(e^{2}a^{2})/(2k.p_{1})\big]k$ ~~&~ $Q_{1}$ &~~ $\vec{q}_{1}$ &~~ $m_{e}^{*}=\sqrt{m_{e}^{2}-e^{2}a^{2}}$  \\  \hline
Scattered &~~ $q_{2}=p_{2}-\big[(e^{2}a^{2})/(2k.p_{2}\big]k$ ~~&~ $Q_{2}$ &~~ $\vec{q}_{2}$ &~~ $m_{e}^{*}=\sqrt{m_{e}^{2}-e^{2}a^{2}}$   \\ \hline
%Nucléon initial &~~ $p$  ~~& $E_{p}$ &~~ $\vec{p}$   \\ \hline
%Nucléon final &~~ $p'$ ~~& $E_{p'}$ &~~ $\vec{p'}$   \\ \hline
%Transfert énergie-impulsion &~ $q$ ~~& $-$ &~ $\vec{q}$   \\ \hline
 \toprule
\end{tabular}
\end{center}
\caption{Different parameters of the initial and final electrons in the electron-nucleon diffusion inside a laser field.}\label{table2}
\end{table}
When a laser field is applied to an electron (incident and scattered), we replace its free-wave functions in the expression of the cross-section by Dirac-Volkov functions $\psi_{e_{1}}$ and $\psi_{e_{2}}$ \cite{Volkov}, which are as follows:
\begin{eqnarray}\label{eq15}
\psi_{e_{1}}(x)=[1+\dfrac{e\slashed{A}\slashed{k}}{2(k.p_{1})}]\dfrac{u_{e}(p_{1},s_{1})}{\sqrt{2Q_{1}V}}e^{iS(q_{1},x)}\qquad\text{and}\qquad\psi_{e_{2}}(x)=[1+\dfrac{e\slashed{A}\slashed{k}}{2(k.p_{2})}]\dfrac{u_{e}(p_{2})}{\sqrt{2Q_{2}V}}e^{iS(q_{2},x)}.
\end{eqnarray}
In this case, the scattering matrix element that corresponds to the virtual photon $\gamma$ Feynman diagram will be modified, and it has the following form:
\begin{eqnarray}\label{eq16}
S_{fi}^{\gamma}=\int d^{4}xd^{4}yJ^{\gamma_{e}}_{\mu}(x)G^{\gamma\mu\nu}(x-y)J^{\gamma_{p}}_{\nu}(y).
\end{eqnarray}
Because it is neutral, the electromagnetic Feynman propagator $G_{\gamma}^{\mu\nu}(x-y)$ keeps the same expression as in the absence of the laser field. In addition, since we have embedded only the final and initial electrons in the laser field, the current $J^{\gamma_{p}}_{\nu}(y)$ remains the same as in the absence of the laser field.
\begin{eqnarray}\label{eq17-18}
G_{\gamma}^{\mu\nu}(x-y)&=&\int \dfrac{d^{4}q}{(2\pi)^{4}}\dfrac{e^{-iq(x-y)}}{q^{2}}[-ig^{\mu\nu}+i\dfrac{q^{\mu}q^{\nu}}{q^{2}}],\\
J^{\gamma_{p}}_{\nu}(y)&=&\phi^{*}_{p^{\prime}}(y)\dfrac{e}{(2\pi)^{3}}\left\langle p_{2}| J_{em_{\nu}}(0)| p_{1}\right\rangle \phi_{p}(y), 
\end{eqnarray}
where $\phi_{p}(y)$ and $\phi_{p'}(y)$ are the wave functions that represent the nucleon in the initial and final states, respectively.
\begin{eqnarray}\label{eq19}
\phi_{p}(y)=\dfrac{1}{\sqrt{2E_{p}V}}e^{-ip.y}\qquad\text{and}\qquad\phi_{p'}(y)=\dfrac{1}{\sqrt{2E_{p'}V}}e^{-ip'.y}.
\end{eqnarray}
 In the current of dressed electrons $J^{\gamma_{e}}_{\mu}(x)$, we replace the free-Dirac functions with Dirac-Volkov states such that:
\begin{eqnarray}\label{eq20}
J^{\gamma_{e}}_{\mu}(x)&=&\bar{\psi}_{e_{2}}(x)[-ie\gamma_{\mu}]\psi_{e_{1}}(x).
\end{eqnarray}
By inserting the Volkov states (\ref{eq15}), the current $J^{\gamma_{e}}_{\mu}(x)$ in equation (\ref{eq20}) will be as follows:
\begin{eqnarray}\label{eq21}
J^{\gamma_{e}}_{\mu}(x)=\dfrac{-ie}{\sqrt{4Q_{1}Q_{2}V^{2}}}e^{i[S(q_{1},x)-S(q_{2},x)}\bar{u}_{e}(p_{2},s_{2})\left[1+\dfrac{e\slashed{A}\slashed{k}}{2(k.p_{2})}\right]\gamma_{\mu}\left[1+\dfrac{e\slashed{k}\slashed{A}}{2(k.p_{1})}\right]u_{e}(p_{1},s_{1}).
\end{eqnarray}
After expanding the equation (\ref{eq16}), the scattering matrix element $S_{fi}^{\gamma}$ becomes:
\begin{eqnarray}\label{eq22}
S_{fi}^{\gamma}&=&\dfrac{-e^{2}Z}{4V^{2}\sqrt{Q_{1}Q_{2}E_{p_{1}}E_{p_{2}}}}\dfrac{1}{(2\pi)^{6}}\int d^{4}x\frac{e^{-i(p_{1}-p_{2})x}}{(p-p')^{2}}(p+p')_{\mu}F(q^{2}) \nonumber \\ 
&\times&\bar{u}_{e}(p_{2},s_{2})\left[1+\dfrac{e\slashed{A}\slashed{k}}{2(k.p_{2})}\right]\gamma_{\mu}\left[1+\dfrac{e\slashed{k}\slashed{A}}{2(k.p_{1})}\right]u_{e}(p_{1},s_{1})e^{i[S(q_{1},x)-S(q_{2},x)}.
\end{eqnarray}
For the $Z^{0}$-boson exchange, the scattering matrix element is as follows:
\begin{eqnarray}\label{eq23}
S_{fi}^{Z^{0}}=\int d^{4}xd^{4}yJ^{Z^{0}_{e}}_{\mu}(x)D^{Z^{0}\mu\nu}(x-y)J^{Z^{0}_{p}}_{\nu}(y),
\end{eqnarray}
where the electron current $J^{Z^{0}_{e}}_{\mu}(x)$, the boson propagator $D^{Z^{0}}(x-y)$, and the nucleon current $J^{Z^{0}_{p}}_{\mu}(x)$ are as follows:
\begin{eqnarray}\label{eq24}
J^{Z^{0}_{e}}_{\mu}(x)&=&\bar{\psi}_{e_{2}}(x)\dfrac{-ie}{16\sin^{2}(\theta_{w})\cos^{2}(\theta_{w})}\gamma_{\mu}[(1-4\sin^{2}(\theta_{w}))-\gamma_{5}]\psi_{e_{1}}(x),\\
D^{Z^{0}\mu\nu}(x-y)&=&\int \dfrac{d^{4}q}{(2\pi)^{4}}\dfrac{e^{-iq(x-y)}}{q^{2}-M_{Z^{0}}^{2}}[-ig^{\mu\nu}+i\dfrac{q^{\mu}q^{\nu}}{M_{Z^{0}}^{2}}],\\
J^{Z^{0}_{p}}_{\nu}(y)&=&e\phi^{*}_{p'}(y)\left\langle p'|K_{\nu}(0)| p\right\rangle \phi_{p}(y).
\end{eqnarray} 
By inserting the last three equations into the equation (\ref{eq23}), the scattering matrix element, $S_{fi}^{Z^{0}}$, will be as follows:
\begin{eqnarray}\label{eq25}
S_{fi}^{Z}&=&\dfrac{-e^{2}Z}{4V^{2}\sqrt{Q_{1}Q_{2}E_{p}E_{p'}}}\dfrac{Z}{q^{2}}F(q^{2})(p+p')_{\mu}\dfrac{[(1-4\sin^{2}(\theta_{w}))-N/Z]}{16\sin^{2}(\theta_{w})\cos^{2}(\theta_{w})}\int d^{4}xe^{-i(p-p')x}\nonumber \\
&\times& \dfrac{q^{2}}{q^{2}-M_{Z^{0}}^{2}}\bar{u}_{e}(p_{2},s_{2})\left[1+\dfrac{e\slashed{A}\slashed{k}}{2(k.p_{2})}\right]\gamma_{\mu}[(1-4\sin^{2}(\theta_{w}))-\gamma_{5}]\left[1+\dfrac{e\slashed{k}\slashed{A}}{2(k.p_{1})}\right]u_{e}(p_{1},s_{1})\nonumber\\
&\times&e^{i[S(q_{1},x)-S(q_{2},x)]}.
\end{eqnarray} 
To take into account both $Z^{0}$-boson and $\gamma$ photon exchange in electron-nucleon scattering, we made the summation of the two corresponding scattering matrices as follows:
\begin{eqnarray}\label{eq26}
S_{fi}^{\gamma}+S_{fi}^{Z^{0}}&=&\dfrac{-e^{2}}{4V^{2}\sqrt{Q_{1}Q_{2}E_{p}E_{p'}}}\dfrac{Z}{q^{2}}\textit{v}(q^{2})F(q^{2})(p+p')_{\mu}\int d^{4}xe^{-i(p-p')x}\bar{u}_{e}(p_{2},s_{2})\nonumber \\
&\times &\bigg[1+\dfrac{e\slashed{A}\slashed{k}}{2(k.p_{2})}\bigg]\gamma_{\mu}\bigg[1+a(q^{2})\gamma_{5}\bigg]\bigg[1+\dfrac{e\slashed{k}\slashed{A}}{2(k.p_{1})}\bigg]u_{e}(p_{1},s_{1})e^{i[S(q_{1},x)-S(q_{2},x)]}.
\end{eqnarray}
The equation (\ref{eq26}) is developed using the following three variables  $R=\sqrt{\alpha_{1}^{2}-\alpha_{2}^{2}}$, $\cos(\phi_{0})=\alpha_{1}/R$, and $\sin(\phi_{0})=\alpha_{2}/R$, where $\alpha_{i}=e\big[(a_{i}.p_{1})/(k.p_{1})-(a_{i}.p_{2})/(k.p_{2})\big]$ for $i=\{1,2\}$. Then, we find that:
\begin{eqnarray}\label{eq27}
S_{fi}^{\gamma}+S_{fi}^{Z}&=&\dfrac{-e^{2}}{4V^{2}\sqrt{Q_{1}Q_{2}E_{p}E_{p'}}}\dfrac{Z}{q^{2}}\textit{v}(q^{2})F(q^{2})\sum_{\ell=-\infty}^{\ell=+\infty}\bar{u}_{e}(p_{2},s_{2})\bigg[(p+p')_{\mu}( C_{\mu}^{1}B_{\ell}(R) \nonumber\\
&+&C_{\mu}^{2}B_{1\ell}(R)+C_{\mu}^{3}B_{2\ell}(R))\bigg]u_{e}(p_{1},s_{1})(2\pi)^{4}\delta^{4}(q_{2}-q_{1}-q-\ell k),
\end{eqnarray}
where the quantities $B_{\ell}(R)$, $B_{1\ell}(R)$, and $B_{2\ell}(R)$ are calculated based on the Jacobi-Anger identity \cite{Andrews}, and they are given by:
\begin{align}\label{eqBessel}
\begin{split}
\begin{bmatrix}
B_{\ell}(R)\\
B_{1\ell}(R)\\
B_{2\ell}(R) \end{bmatrix}=\begin{bmatrix}J_{\ell}(R)e^{i\ell\phi_{0}}\\
\big(J_{\ell+1}(R)e^{i(\ell+1)\phi_{0}}+J_{\ell-1}(R)e^{i(\ell-1)\phi_{0}}\big)/2\\
\big(J_{\ell+1}(R)e^{i(\ell+1)\phi_{0}}-J_{\ell-1}(R)e^{i(\ell-1)\phi_{0}}\big)/2i
 \end{bmatrix}.
\end{split}
\end{align}
In equation (\ref{eq27}), the parameters $C_{\mu}^{1}$, $C_{\mu}^{2}$ and $ C_{\mu}^{3}$ are explicitly expressed as follows:
\begin{eqnarray}\label{eq28}
C_{\mu}^{1}&=&\gamma_{\mu}-2C(k)C(k_{2})k_{\mu}\slashed{k}a^{2}+a(q^{2})\gamma_{5},\\
C_{\mu}^{2}&=&C(k)(\gamma_{\mu}+\gamma_{5})\slashed{k}\slashed{a_{1}}+C(k')\slashed{a_{1}}\slashed{k}(\gamma_{\mu}+\gamma_{5}),\\
C_{\mu}^{2}&=&C(k)(\gamma_{\mu}+\gamma_{5})\slashed{k}\slashed{a_{2}}+C(k')\slashed{a_{2}}\slashed{k}(\gamma_{\mu}+\gamma_{5}).
\end{eqnarray}
By inserting the scattering matrix element (\ref{eq27}) in the expression of the differential cross section as in equation (\ref{eq10}), and after some algebraic calculation, we find that the laser-assisted differential cross section is defined as an infinite harmonic summation of order $\ell$ such that:
\begin{eqnarray}\label{eq34}
\dfrac{d\overline{\sigma}}{d\Omega_{f}}=\sum_{\ell=-\infty}^{\ell=+\infty}\dfrac{d\overline{\sigma}^{\ell}}{d\Omega_{f}} &=& \dfrac{Z^{2}e^{4}}{8(2\pi)^{2}m_{p}}\dfrac{|\vec{q_{2}}|}{|\vec{q_{1}}|}\dfrac{|\textit{v}(q^{2})|^{2}| F(q^{2})|^{2}}{| q|^{4}} \sum_{\ell=-\infty}^{\ell=+\infty}\dfrac{|\overline{\mathcal{M}^{\ell}_{fi}}|^{2}}{h'(Q_{2})}, 
\end{eqnarray}
where the $ h'(Q_{2})$ function is the derivative of $h(Q_{2})=(p+q_{1}-q_{2}+\ell k)^{2}-m_{N}^{2}$ with respect to $Q_{2}$.
\section{Results and discussion}\label{Sec2}
After dealing with the theoretical treatment of electron-nucleon scattering in the absence and presence of a laser field, we will explore in this part the numerical analysis of the obtained results. 
As mentioned in the theoretical framework, the DCS of the electron-nucleon diffusion has two factors: The number of neutrons $N$ and the number of protons $Z$. 
To be more precise, for the electron-proton (e-p) scattering, we will consider the scenario where $Z=1$ and $N=0$, while $Z=0$ and $N=1$ for the case of electron-neutron (e-n) scattering.  
We use spherical geometry for both the absence and presence of a laser field such that $\theta_{i}$ and $\phi_{i}$ are the spherical coordinates of the incoming electron, while $\theta_{f}$ and $\phi_{f}$ are the spherical angles representing the scattered electron. 
Throughout this work, we choose $\theta_{i}=\phi_{i}=\phi_{f}=0^{\circ}$, and the direction of the vector wave is taken along the (oz) axis ($\vec{k}=k.\vec{e}_{z}$).
The traces that appear in $ |\overline{\mathcal{M}_{fi}}|^{2} $ and $ |\overline{\mathcal{M}^{\ell}_{fi}}|^{2} $ are computed by the symbolic-algebra program FeynCalc \cite{Feyncalc}, whereas the numerical evolution of the DCS is performed by using the Mathematica programming language. 
We begin our discussion by analyzing how the electromagnetic field's strength and its frequency affect the number of transferred photons $\ell$.
\begin{figure}[H]
\centering
    \includegraphics[scale=0.48]{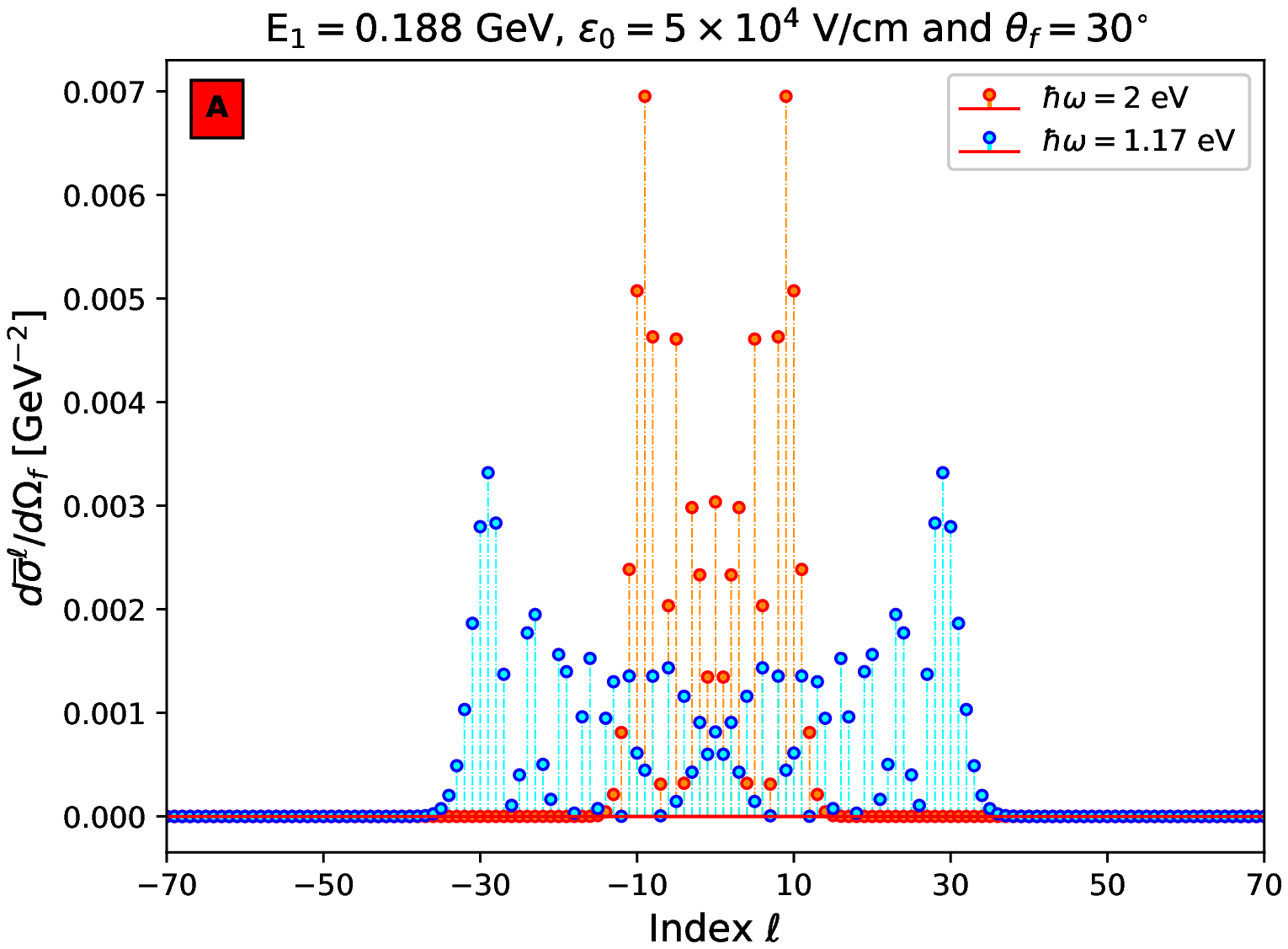}% \hspace*{0.1cm}
    \includegraphics[scale=0.48]{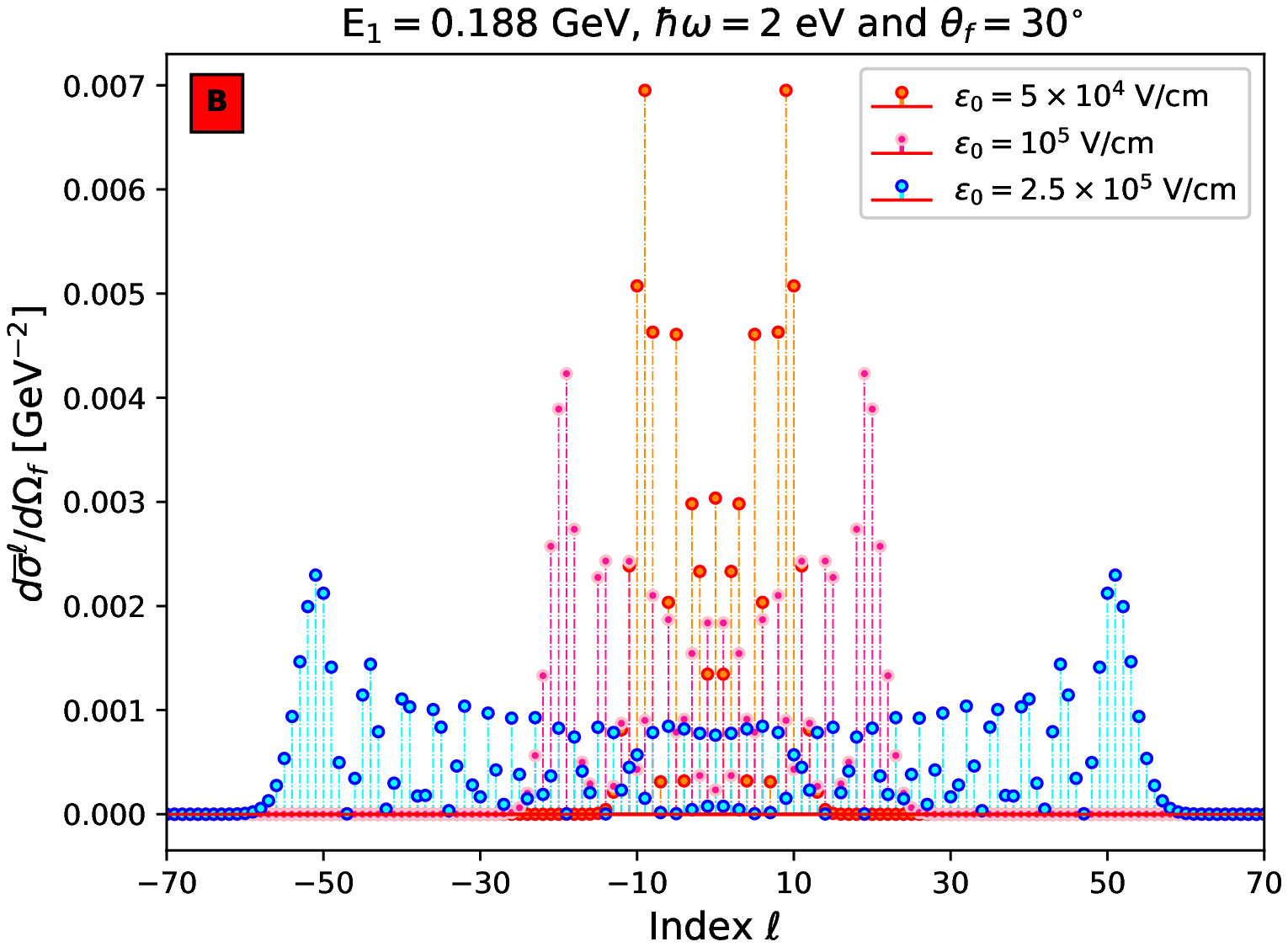} \par
 \hspace*{0.15cm}   \includegraphics[scale=0.48]{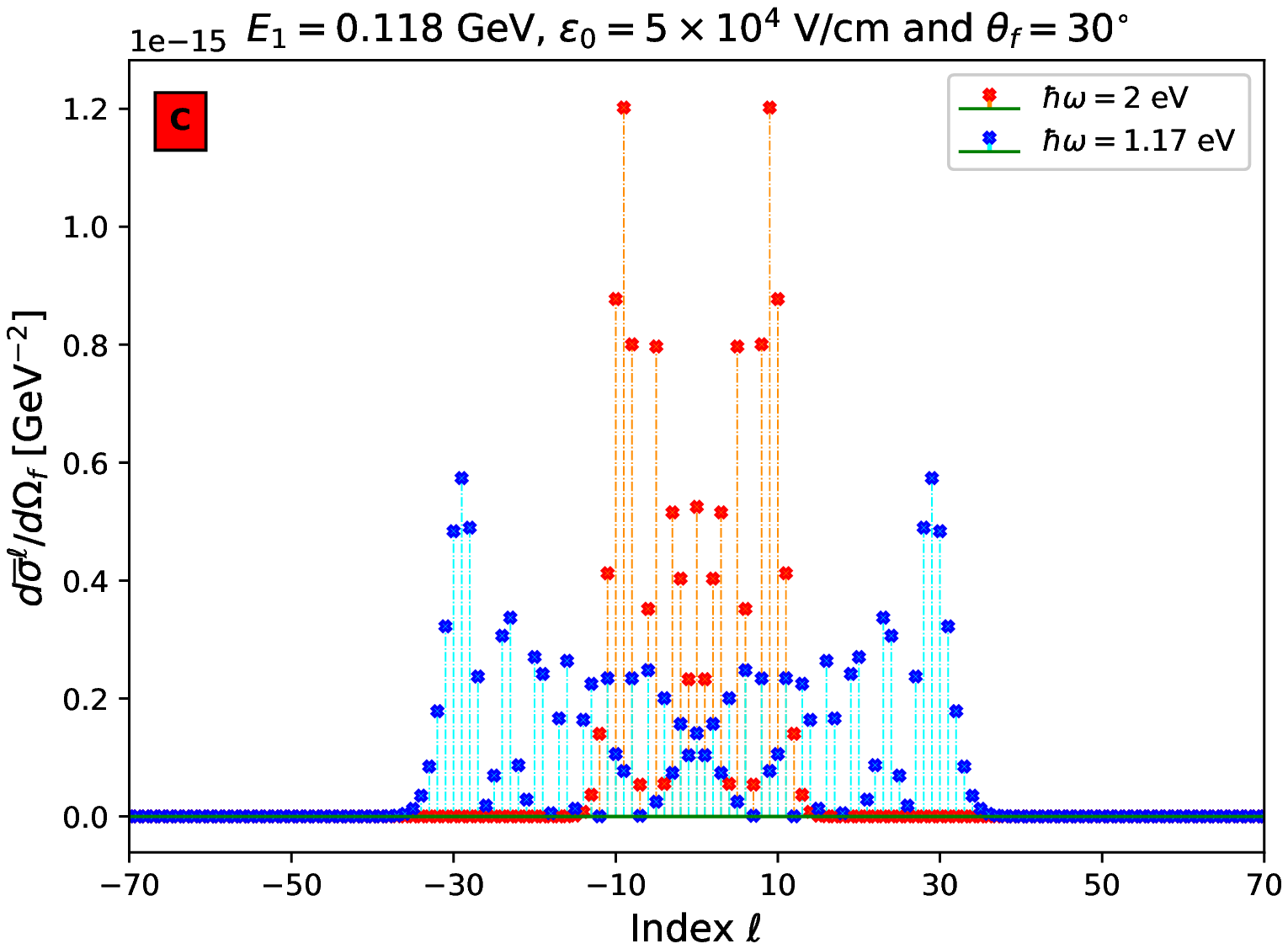} 
    \includegraphics[scale=0.48]{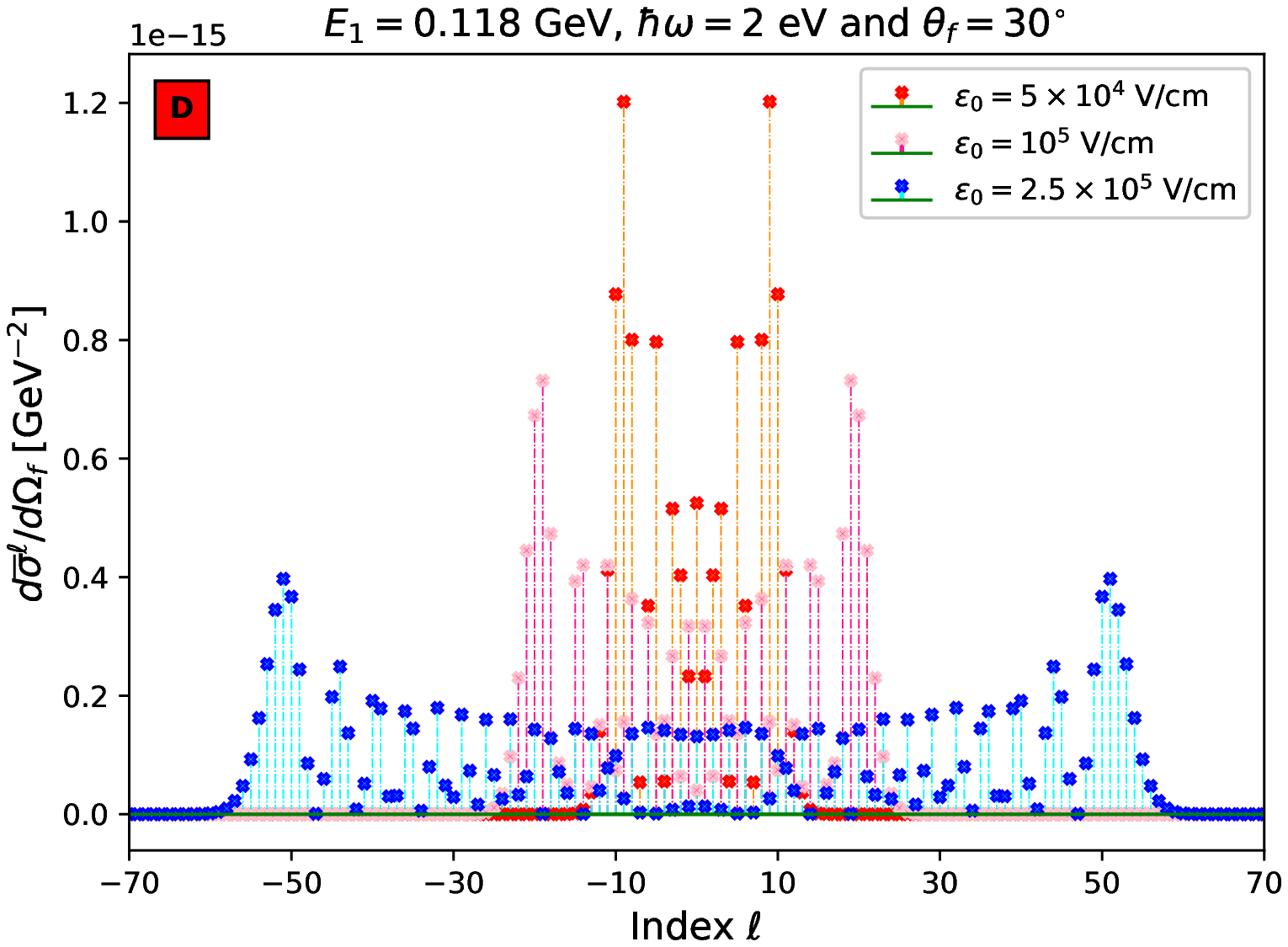}
  \caption{Variation of $d\bar{\sigma}^{\ell}/d\Omega_{f}$ as a function of the number of photons exchanged $\ell$
for different frequencies $\hbar\omega$ (\ref{Figure:1}-A and \ref{Figure:1}-B)/laser strengths $\varepsilon_{0}$ (\ref{Figure:1}-C and \ref{Figure:1}-D). The incident electron energy and the scattering angle are $E_{1}=0.188\,GeV$ and $\theta_{f}=30^{\circ}$, respectively. Upper panel (\ref{Figure:1}-A and \ref{Figure:1}-B): electron-proton scattering ; lower panel (\ref{Figure:1}-C and \ref{Figure:1}-D): electron-neutron scattering.} \label{Figure:1}
\end{figure}
The dependence of the differential cross section $d\bar{\sigma}^{\ell}/d\Omega_{f}$ on the number of transferred photons $\ell$ for various values of $\hbar\omega$ and $\varepsilon_{0}$ is shown in figure \ref{Figure:1} where $E_{k}=0.188\,GeV$ and $\theta_{f}=30^{\circ}$. 
The total DCS is the summation of the partial DCS over $\ell$.
According to figure \ref{Figure:1}, when a circularly polarized laser field is applied, its frequency and strength become important in determining the number of photons that may be exchanged during the scattering process. 
In fact, the function $\delta^{4}(q_{2}-q_{1}-p+p'-\ell k)$, which assures the conservation of the four-momentum (Eq.(\ref{eq27})), introduces the interaction of the physical system with $\ell$ photons of wave four-vector $k$.
This interaction leads to the interpretation of the index $\ell$ as a number of photons (See p. 451 in reference \cite{Landau}). 
The value of $\ell$ determines how many photons are transferred, and its sign indicates whether these photons are emitted ($\ell < 0$) or absorbed ($\ell > 0$). 
In figure \ref{Figure:1}-A, we compare $d\bar{\sigma}^{\ell}/d\Omega_{f}$ as a function of $\ell$ in the case of electron-proton scattering for the Nd$:$YAG laser field with frequency $ \hbar\omega=1.17 \,eV$ with that in the case of a pulsed He$:$Ne laser field with frequency $\hbar\omega=2 \,eV$.
Indeed, we observe that as far as the laser field frequency decreases, $d\sigma^{\ell}/d\Omega_{f}$ also decreases, while the number of photons that may be interchanged increases.
In figure \ref{Figure:1}-B, the frequency is fixed at a value $\hbar\omega=2\,eV$ and the strength of the laser field $\varepsilon_{0}$ takes the values $5\times10^{4}\,V/cm$, $10^{5}\,V/cm$ and $2.5\times10^{5}\,V/cm$.
We still see that $d\bar{\sigma}^{\ell}/d\Omega_{f}$ decreases and the number $\ell$ increases as in figure \ref{Figure:1}-A, but this time by increasing the laser field strength $\varepsilon_{0}$.
In these two figures (\ref{Figure:1}-A, \ref{Figure:1}-B), the quantity $d\bar{\sigma}^{\ell}/d\Omega_{f}$ is symmetrical with respect to $\ell=0$, and this symmetry is due to the property of the ordinary Bessel functions given by the equation (\ref{eqBessel}). 
Additionally, Similar results have been obtained in the case of electron-neutron (e-n) scattering (Figure \ref{Figure:1}-C and \ref{Figure:1}-D). 
By comparing $d\bar{\sigma}^{\ell}/d\Omega_{f}$ for both (e-p) and (e-n) scattering processes, we see that the behavior of the $d\bar{\sigma}^{\ell}/d\Omega_{f}$ is the same but the scale order is different. 
Besides, the quantity $d\bar{\sigma}^{\ell}/d\Omega_{f}$ in these four figures decreases from a specific values  of $|\ell|$ until it reaches zero.
These specific values of $\ell$ numbers are referred to as the multiphoton process cut-offs, for which the DCS becomes a constant.
In this scenario, the scattering system cannot exchange photons anymore, and the infinite sum of $d\bar{\sigma}^{\ell}/d\Omega_{f}$ in the equation (\ref{eq34}) becomes limited from $-$cutoff to $+$cutoff. 
In the following figure, we show how the differential cross section is impacted by the summation over $\ell$ by raising its value.     
\begin{figure}[H]
\centering
    \includegraphics[scale=0.5]{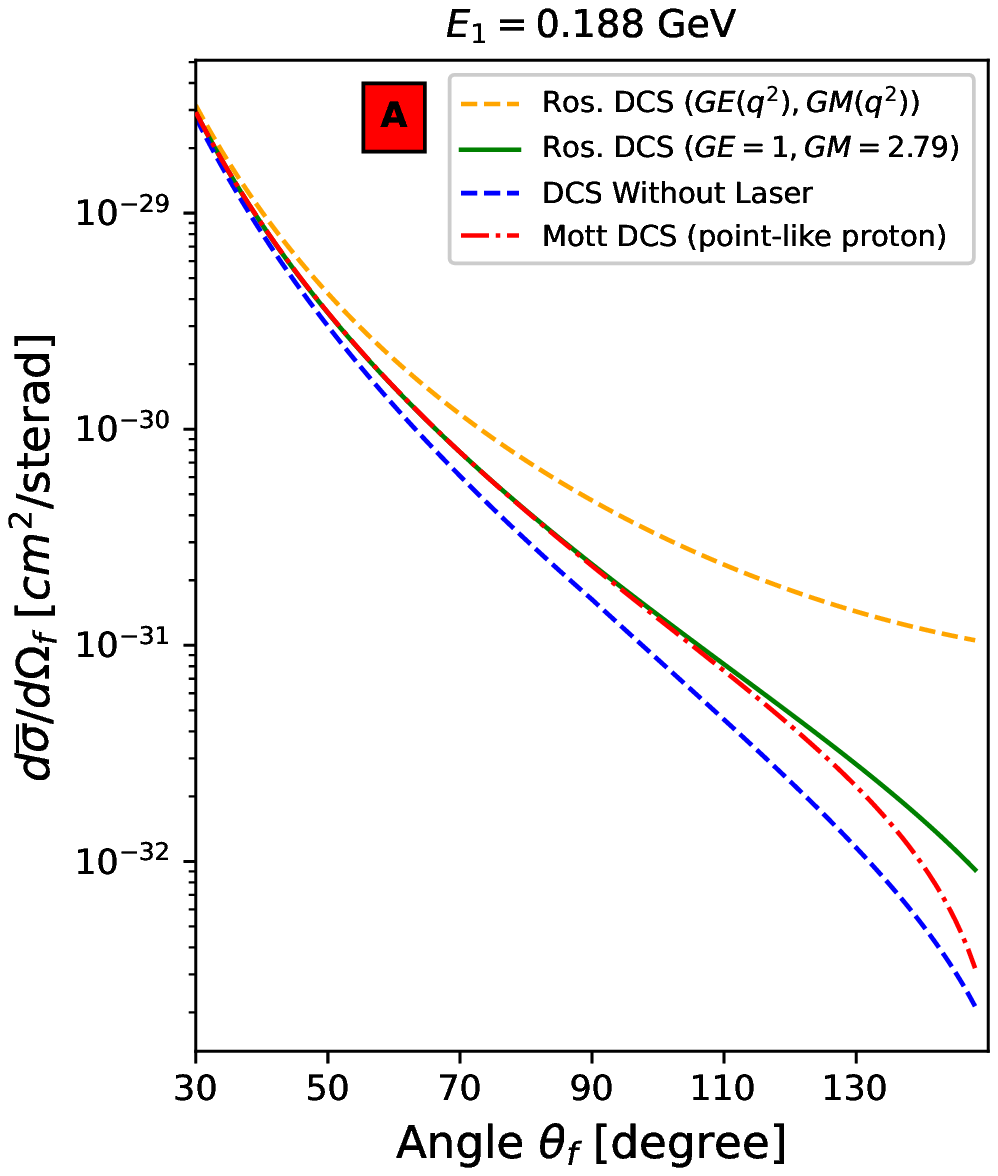}\hspace*{0.25cm}
    \includegraphics[scale=0.5]{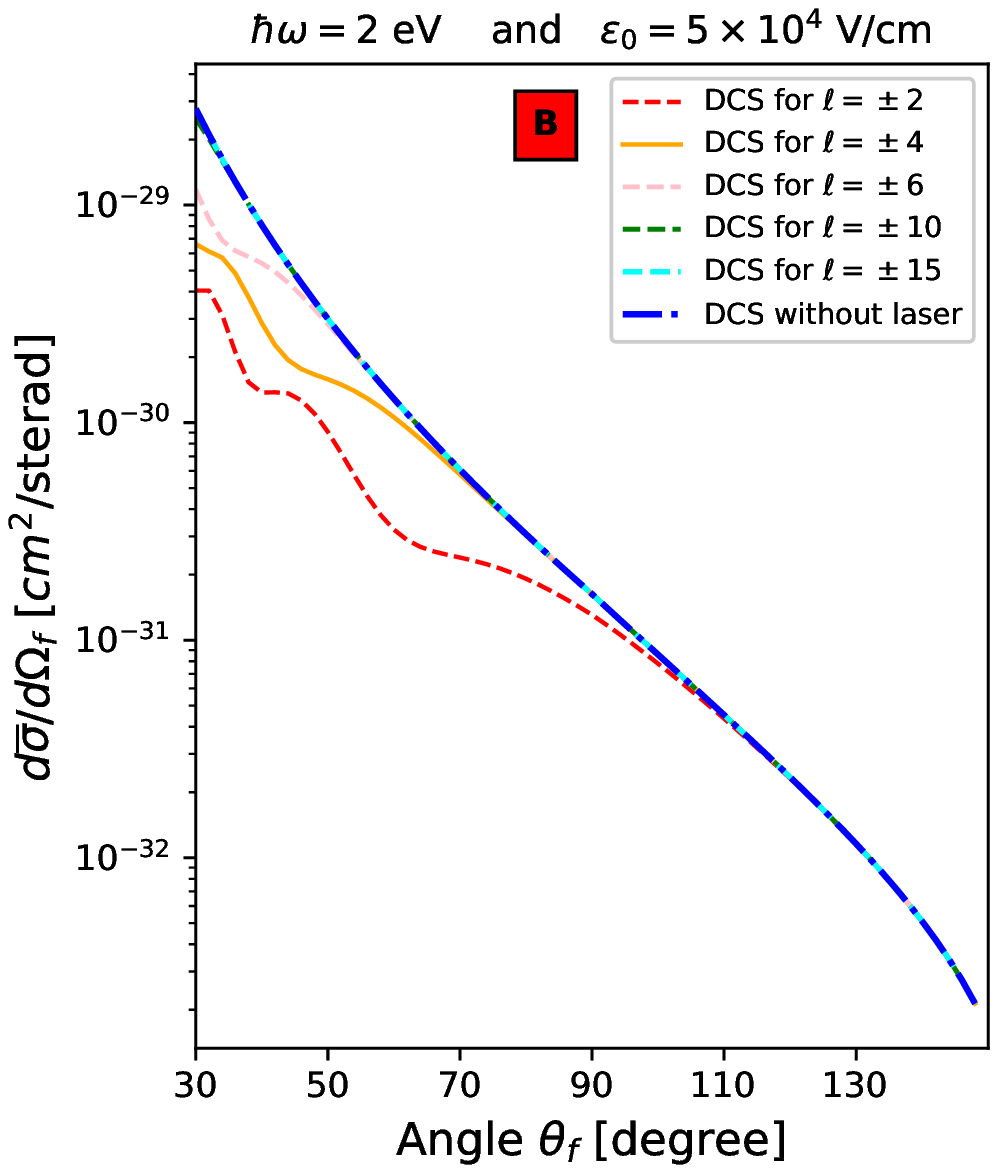}
   \caption{Variation of the differential cross section of the electron-proton scattering as a function of the scattering angle $\theta_{f}$ for the incident electron energy $E_{1}=0.188\,GeV$. Left panel (\ref{Figure:2}-A): The scattering process in the absence of the laser field for different models. Right panel (\ref{Figure:2}-B): The scattering process in the presence of the laser field for $\hbar\omega=2\,eV$, $\varepsilon_{0}=5\times10^{4}\,V/cm$ and for different summations over $\ell$.} \label{Figure:2}
\end{figure}
Before discussing the multiphoton effect on the DCS, we show in figure \ref{Figure:2}-A the laser-free DCS as a function of the electron's scattering angle $\theta_{f}$. Four curves are presented in figure \ref{Figure:2}-A: The first curve in blue represents the (e-p) scattering process in which the proton is regarded as a spinless particle, whereas the red curve represents the case where the proton is a point particle (Mott scattering). The green and orange curves show the Rosenbluth formula \cite{Rosenbluth}, which includes the electric GE and magnetic GM form factors of the proton.
The  $ G_{E} $ and $ G_{M} $ form factors are specified in the orange curve using the usual dipole parametrization, while they are defined as $ G_{E}=1 $ and $ G_{M}=2.79 $ in the green curve. 
In this figure, we can observe that the scattering process represented with a linear electric form factor in $q^{2}$ (see equation (\ref{eq13})) agrees with the other models for small values of $\theta_{f}$, and it starts to differ from the Rosenbluth model (orange curve) at big scattering angles. 
We now go to figure \ref{Figure:2}-B in which we investigated the (e-p) scattering process by using a pulsed laser field He$:$Ne for various summations over $\ell$ with $\varepsilon_{0}=5\times10^{4}\,V/cm$ and $E_{1}=0.188\,GeV$.
For a number of photons exchanged $|\ell|<15$, the DCS in the presence of the laser field is still not confused with the laser-free DCS. 
This result is explained by the fact that the value of the cut-off that corresponds to this laser field strength has not yet been attained. 
However, when the number of transferred photons reaches $|\ell|=15$, the multiphoton cut-off of this process is exceeded and the laser-assisted DCS will be equal to the DCS in the absence of the laser field.
For the (e-n) scattering, we assume that the DCS of the laser field multiphoton interaction with the scattering system is similar to that of the (e-p) scattering, with a difference in the order of magnitude. 
This is due to the fact that we have considered the nucleon as a free particle, and only the electron which is embedded in the laser field.
This characteristic of multiphoton interaction with the laser of circular polarization is frequently found not only in the scattering processes \cite{Attaourti,Manaut3,Manaut4,Ouali1,Ouhammou1,ElAsri,Ouali2,Ouhammou2,Mekaoui} but also in laser-assisted decays \cite{Mouslih1,Jakha1,Jakha2,Mouslih2,Mouslih3,Baouahi1,Baouahi2}. 
To compare the DCS of the (e-p) scattering with that of (e-n) scattering, we exhibit several variations of the DCS in the following figure by varying the number of exchanged photons and the strength of the laser field.
\begin{figure}[H]
\centering
    \includegraphics[scale=0.48]{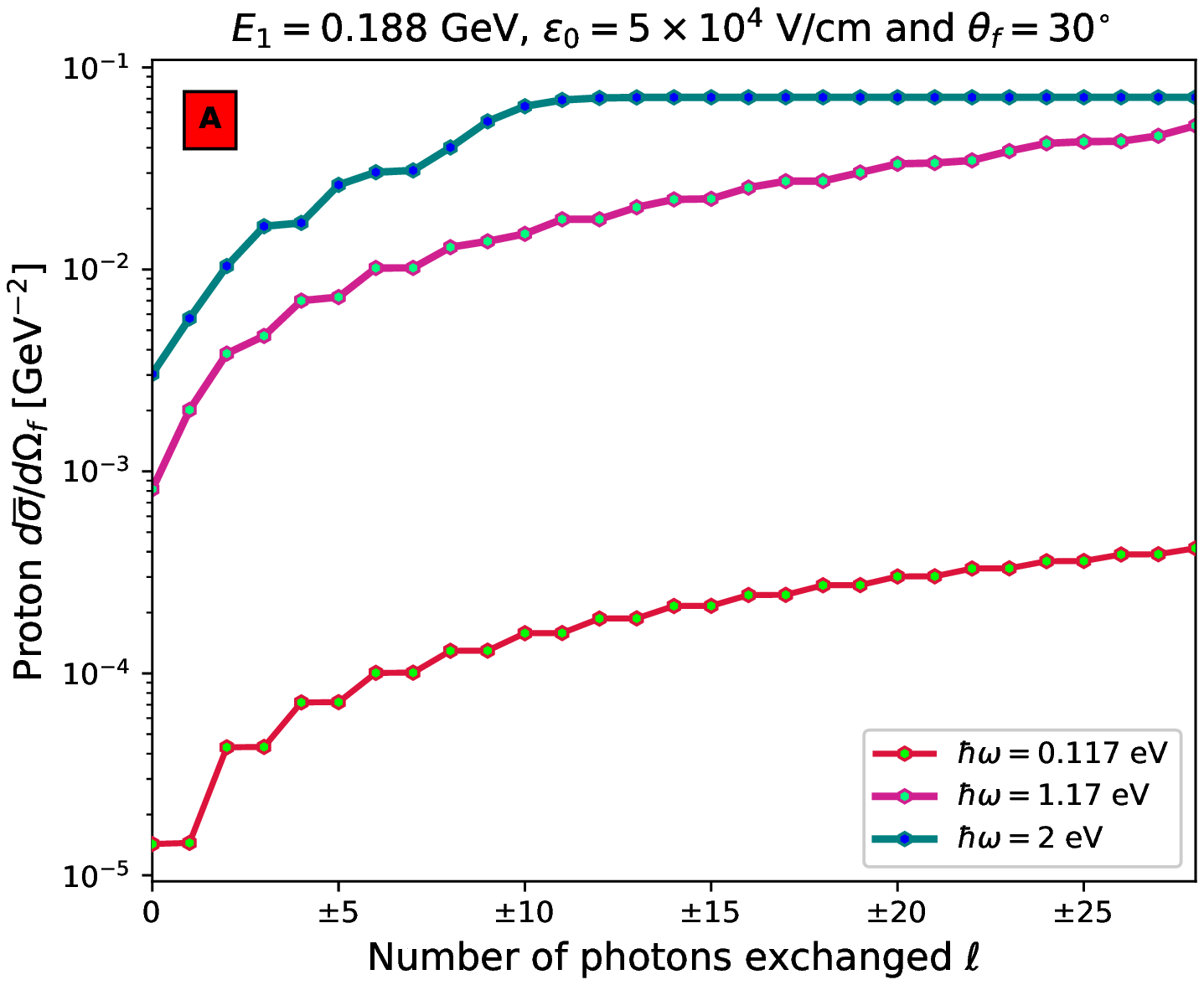} \hspace*{0.1cm}
    \includegraphics[scale=0.48]{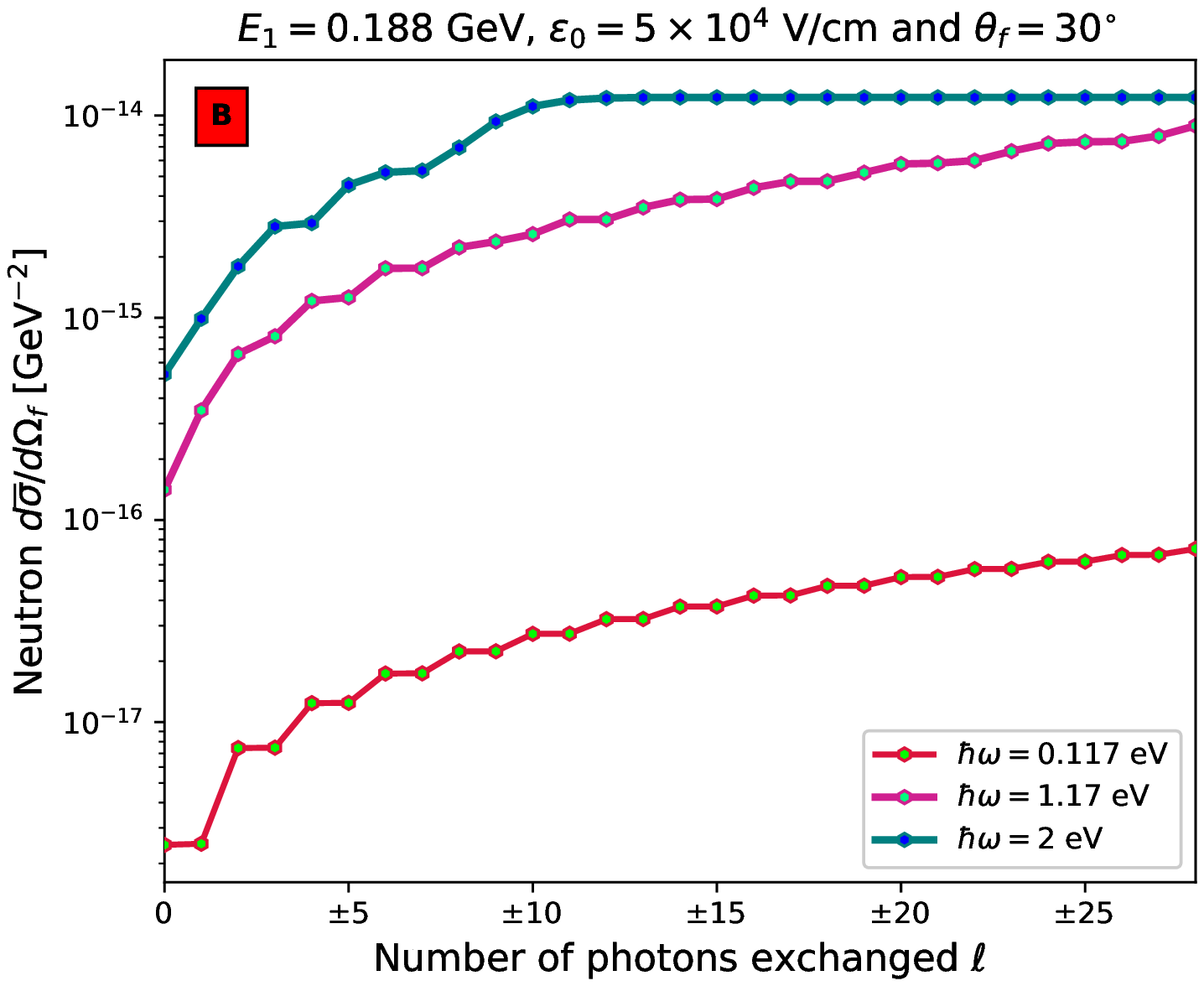} \par
    \includegraphics[scale=0.48]{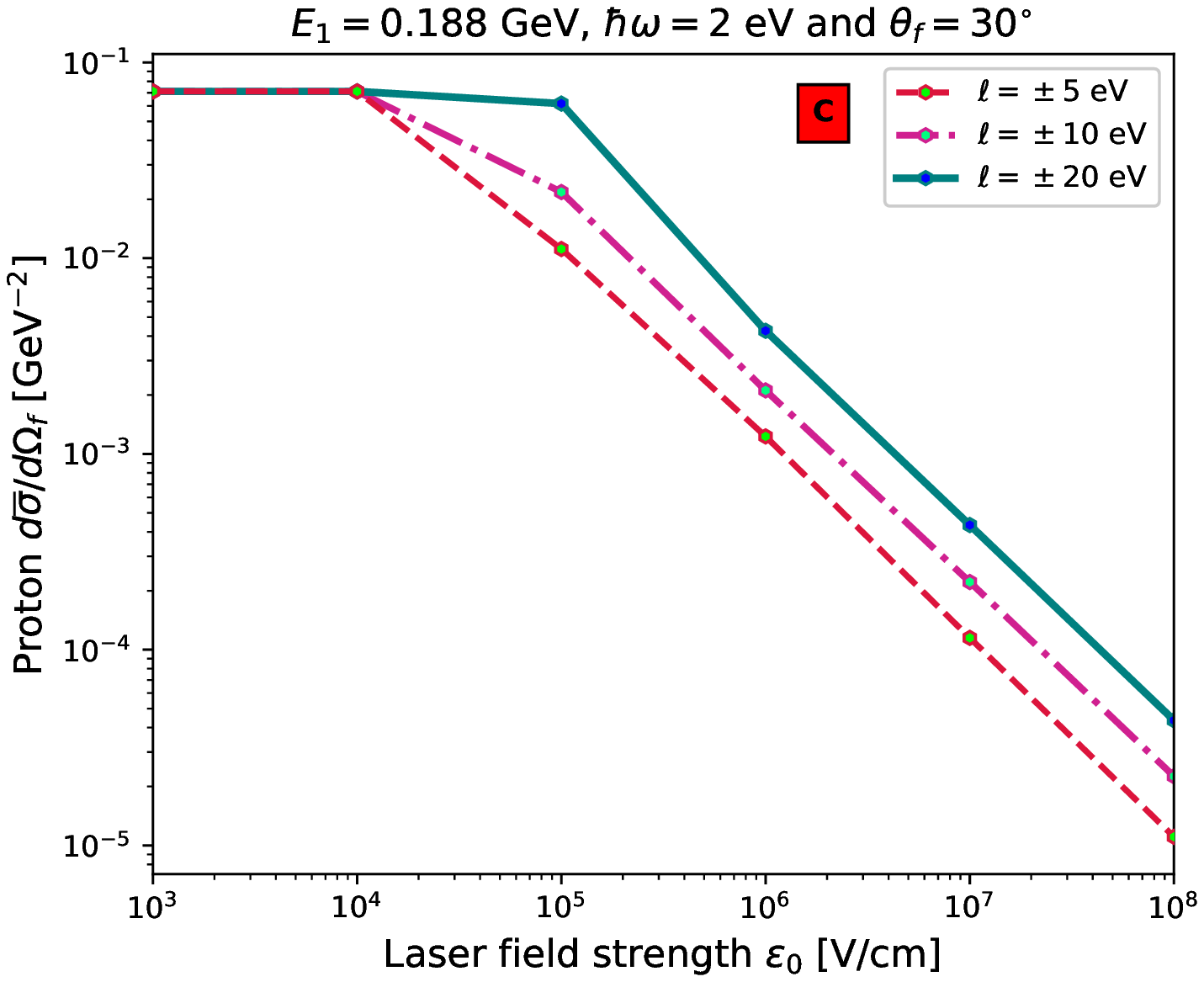} \hspace*{0.1cm}
    \includegraphics[scale=0.48]{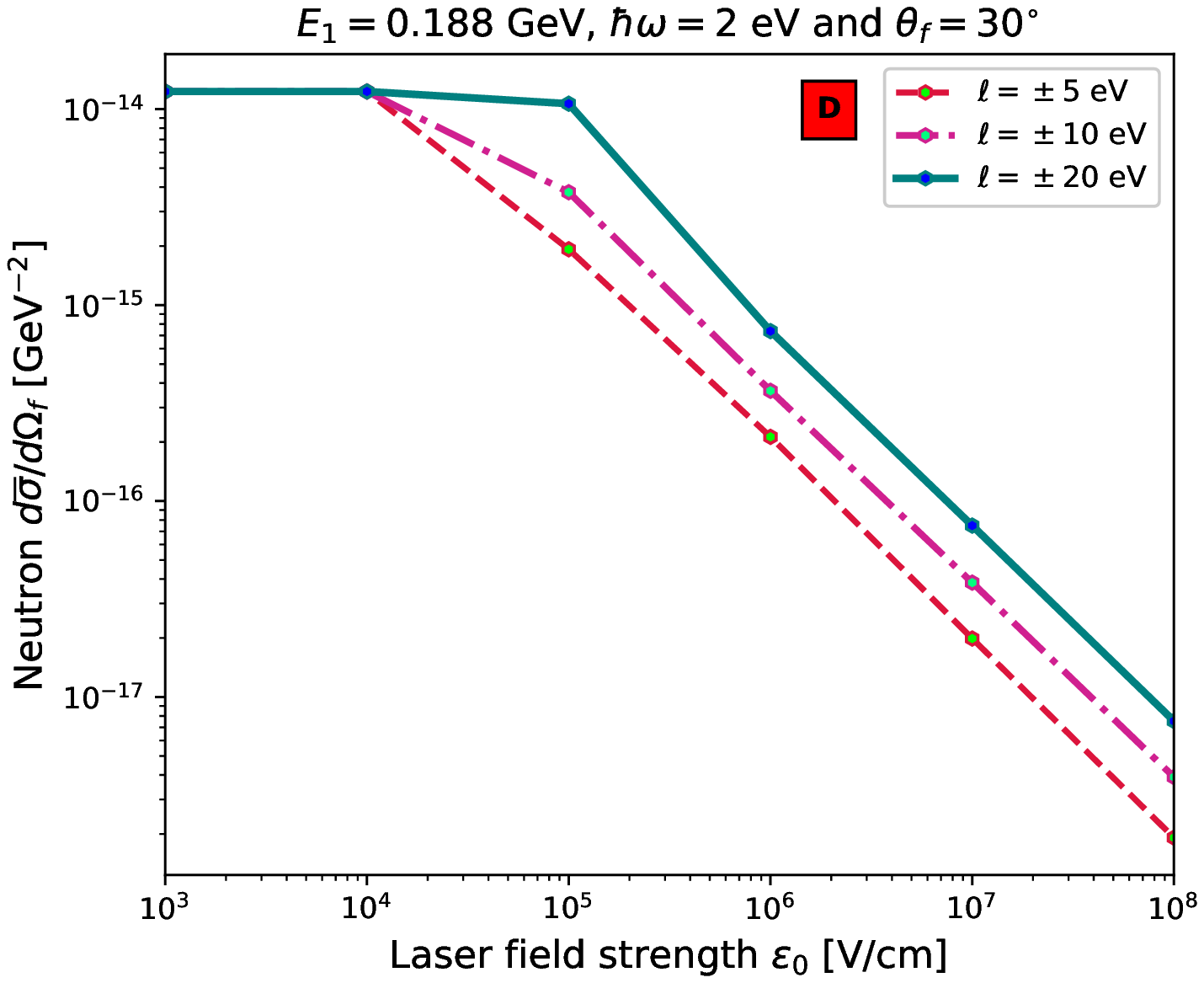}
  \caption{The upper panel: Variation of the differential cross section with respect to the number of photons exchanged $\ell$ for different frequencies by taking $ E_{1}=0.188\,GeV $, $ \theta_{f}=30^{\circ} $ and $ \varepsilon_{0}=5\times10^{4}\,V/cm $.
The lower panel: Variation of the differential cross section as a function of $ \varepsilon_{0} $ for various summations over $\ell$ with $ E_{1}=0.188\,GeV $, $ \theta_{f}=30^{\circ} $ and $\hbar\omega=2\,eV$.
Left column (\ref{Figure:3}-A and \ref{Figure:3}-C): electron-proton scattering. 
Right column (\ref{Figure:3}-B and \ref{Figure:3}-D): electron-neutron scattering.}\label{Figure:3}
\end{figure}
We present in the two figures \ref{Figure:3}-A and \ref{Figure:3}-B the variation of $d\bar{\sigma}/d\Omega_{f}$ with respect to the possible number of photons $\ell$ to be exchanged for (e-p) and (e-n) scattering, respectively.
Different types of electromagnetic field are applied: A pulsed CO$_{2}$ laser field with frequency $\hbar\omega=0.117\,eV$, a pulsed Nd$:$YAG laser field ($\hbar\omega=1.17\,eV$), and a pulsed He$:$Ne laser field ($\hbar\omega=2\,eV$). We mention that the energy of the incident electron, the value of the scattering angle and the laser strength are successively $E_{1}=0.188\,GeV$, $ \theta_{f}=30^{\circ} $ and $\varepsilon_{0}=5\times10^{4}\,V/cm$.
For the pulsed He$:$Ne laser field, we notice that the DCS increases by increasing the number of photons exchanged $\ell$, and from $\ell = \pm10$, the differential cross section becomes constant. 
For the other field frequencies, this phenomenon is possible only for a number of photons $\ell$ higher than $\pm30$, and this result can be seen for the (e-n) scattering as well as for the (e-p) scattering process.
To interpret this phenomenon, we say that an electron dressed by a pulsed laser field He$:$Ne with a strength $\varepsilon_{0}=5\times10^{4}\,V/cm$ can not exceed $20$ photons exchanged ($-10\leqslant\ell\leqslant10$) during the multi-photon scattering process. 
We plot $d\bar{\sigma}/d\Omega_{f}$ for the (e-p) scattering (\ref{Figure:3}-C) and the (e-n) scattering (\ref{Figure:3}-D) by varying the laser field strength from $10^{3}\,V/cm$ to $10^{8}\,V/cm$. We keep the same incident electron energy $E_{1}=0.188\,GeV$ and the scattering angle $\theta_{f}=30^{\circ}$, and we fix the frequency at $\hbar\omega=2\,eV$.
We notice that the DCS begins to decrease from $\varepsilon_{0}=10^{4}\,V/cm$ for different summations over $\ell$ from $-i$ to $+i$, such that $i=\{5, 10\}$. When $i$ is equal to $20$, the decrease becomes apparent from $\varepsilon_{0}=10^{5}\,V/cm$. 
We deduce from figure \ref{Figure:3} that, for both (e-p) scattering and (e-n) scattering, the DCS increases by enlarging the range of photons exchanged $\ell$ regardless of the laser frequency (figures \ref{Figure:3}-A and \ref{Figure:3}-B), while it decreases with the increasing laser field strengths (\ref{Figure:3}-C and \ref{Figure:3}-D).
From figure \ref{Figure:1}-B and for large values of $\varepsilon_{0}$ it is clear that the quantity $d\bar{\sigma}/d\Omega_{f}$ requires much larger number of photons than $20$.
To know more about the variation of the DCS when the laser strength increases, we will plot in \ref{Figure:4}-B its simultaneous dependence on both $\varepsilon_{0}$ and $\ell$.
\begin{figure}[H]
\centering
    \includegraphics[scale=0.5]{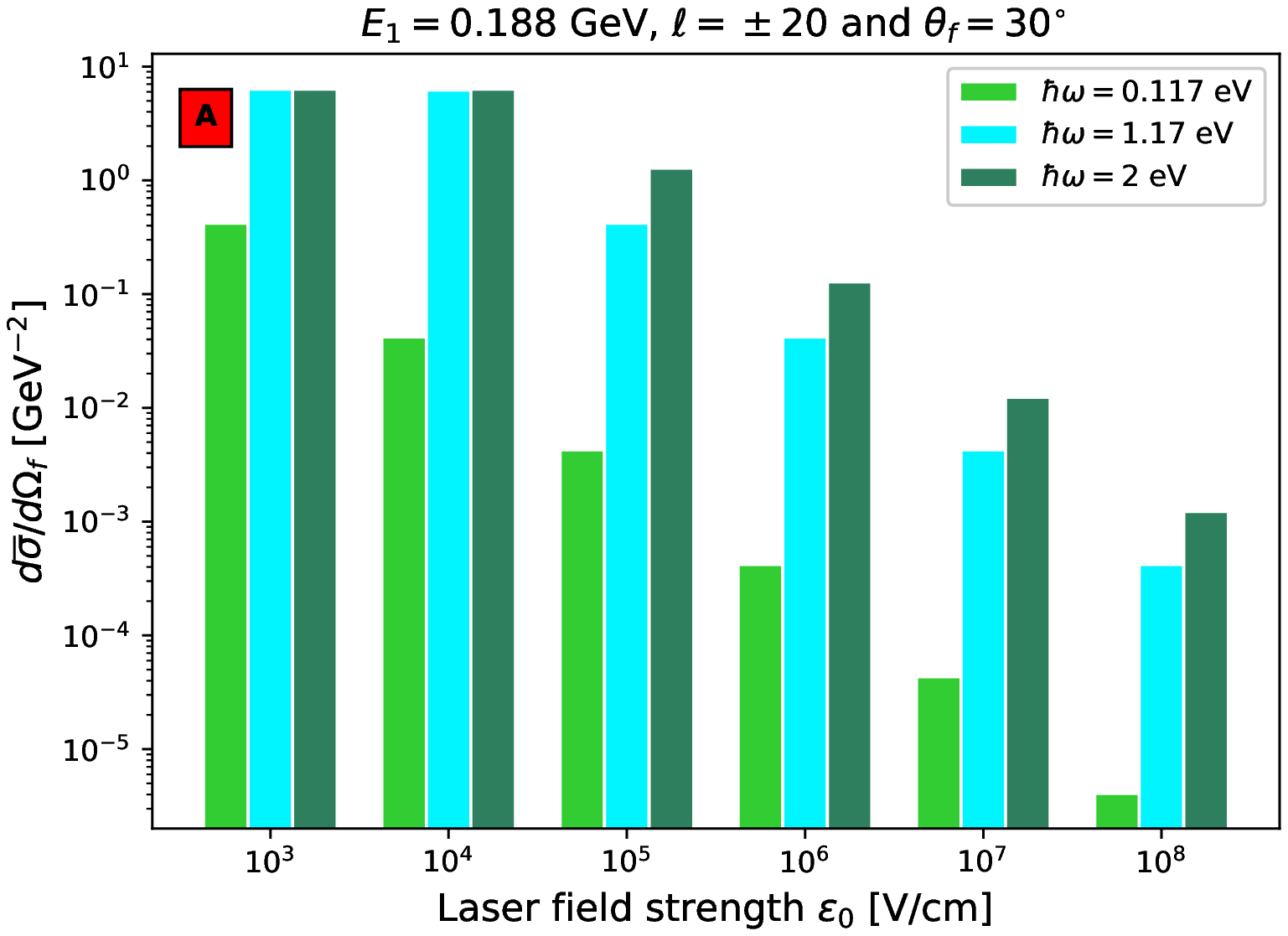}\hspace*{0.25cm}
    \includegraphics[scale=0.5]{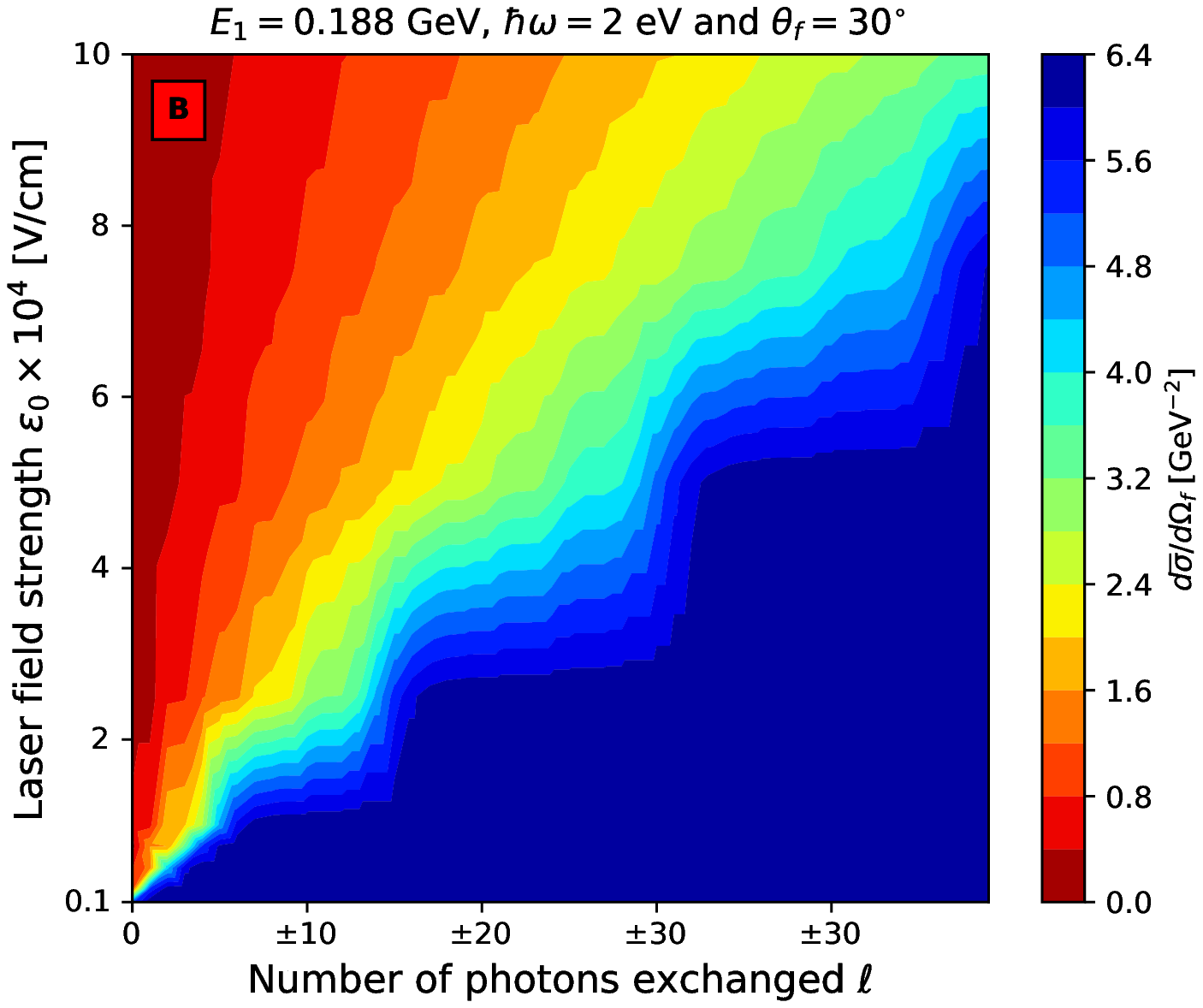}
   \caption{Left panel (\ref{Figure:4}-A): Differential cross section of electron-proton scattering as a function of the laser field strength for various $\hbar\omega$ values and by summing over $ \ell $ from $-20$ to $20$.
Right panel (\ref{Figure:4}-B): Variation of the differential cross section of electron-proton scattering for different values of the laser field strength and for different summation over $ \ell $ with $ \hbar\omega=2\,eV $.
The incident electron energy $E_{1}$ and the scattering angle are $ E_{1}=0.188\,GeV $ and $ \theta_{f}=30^{\circ} $ in both figures.} \label{Figure:4}
\end{figure}
In figure \ref{Figure:4}-A, the histogram shows the DCS of the (e-p) scattering where the electron is dressed by different types of laser fields with different strengths, namely the CO$_{2}$ laser field, the Nd$:$YAG laser field, and the He$:$Ne laser field. 
From this comparison, we notice that increasing the strength of the electromagnetic field leads to the decrease in the DCS of (e-p) scattering even for a fixed number of exchanged photons (a summation over $\ell$ from $-20$ to $20$). 
For the Nd$:$YAG and He$:$Ne laser fields and for $\varepsilon_{0}$ equal to $10^{3}\,V/cm$, the DCS is equal to $6.33363\,GeV^{-2}$, and this value corresponds to the DCS  in the absence of the laser field. 
However, the DCS takes the value $0.41901\,GeV^{-2}$ for the pulsed CO$_{2}$ laser at $\varepsilon_{0}= 10^{3}\,V/cm$.
This result is due to the fact that a pulsed CO$_{2}$ laser field is rich in photons even at small laser field strengths.
We now turn to figure \ref{Figure:4}-B which illustrates the DCS of the (e-p) scattering by varying the number of photons exchanged $\ell$ and the laser field strength $\varepsilon_{0}$ from $10^{3}\,V/cm$ to $10^{5}\,V/cm$.
In this case, the electron in the initial state has the energy $E_{1}=0.188\,GeV$, and the scattering angle is $\theta_{f}=30^{\circ}$. 
We remark that the differential cross section becomes maximal for low values of laser strength and for high values of $\ell$ (the blue area). 
This maximum value represents also the DCS of the (e-p) scattering process in the absence of the laser field, i.e., the laser field no longer has any effect on the scattering process even if we sum over $\ell$ from $-\infty$ to $+\infty$.
On the contrary, the DCS takes small values for high laser field strengths and small numbers of exchanged photons $\ell$ number (red area), and this fact makes the process less likely to be detected. 
From these results, we can say that the (e-p) multiphoton scattering process can be important in all areas of \ref{Figure:4}-B except the blue and red ones which are excluded.
Next, we will move to the last results to see the effect of the laser field as well as the impact of the incident energy $E_{1}$ of the electron on the form factor given by equation (\ref{eq13}) for both (e-p) and (e-n) scattering processes.
\begin{figure}[H]
\centering
    \includegraphics[scale=0.5]{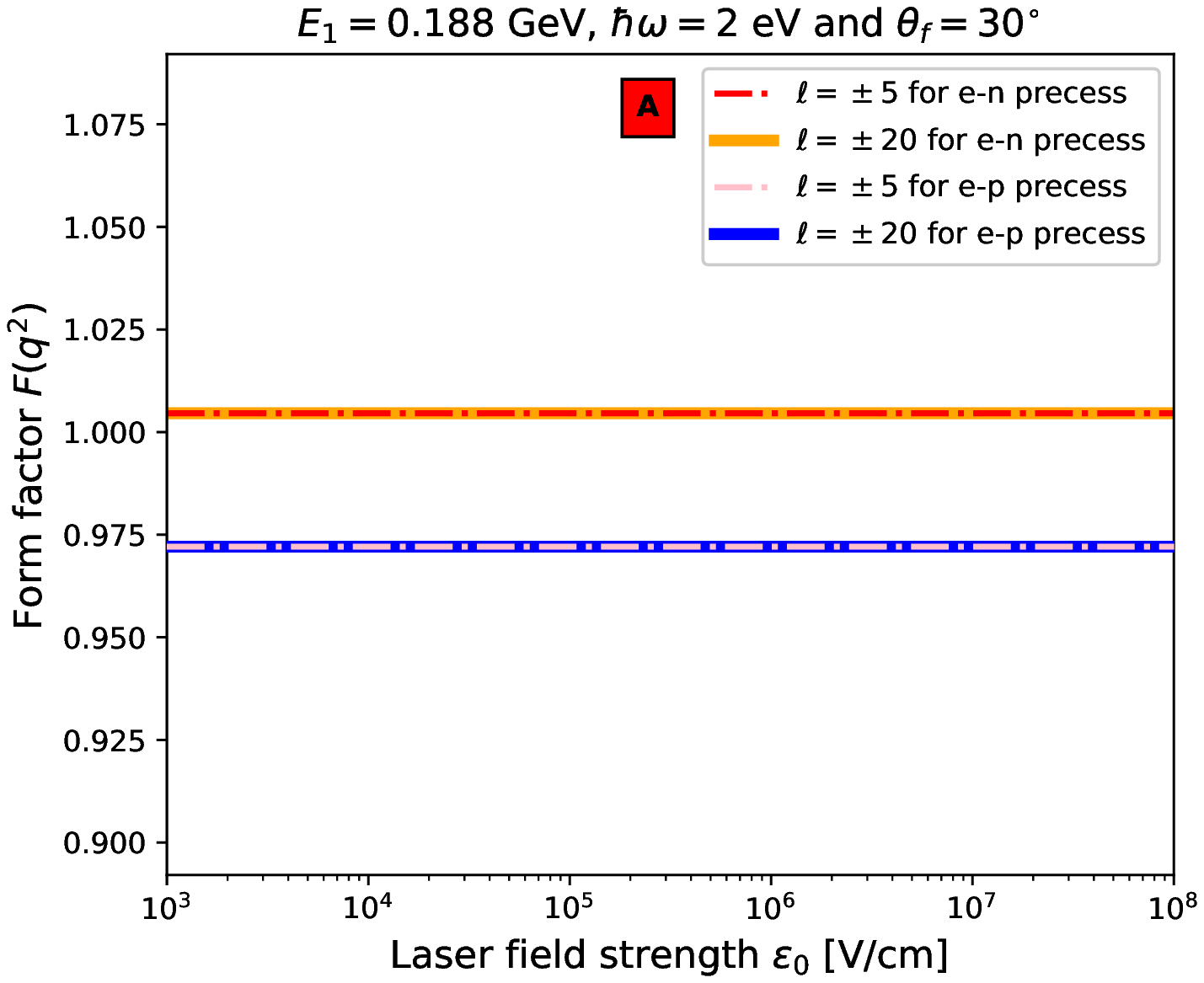}\hspace*{0.25cm}
    \includegraphics[scale=0.5]{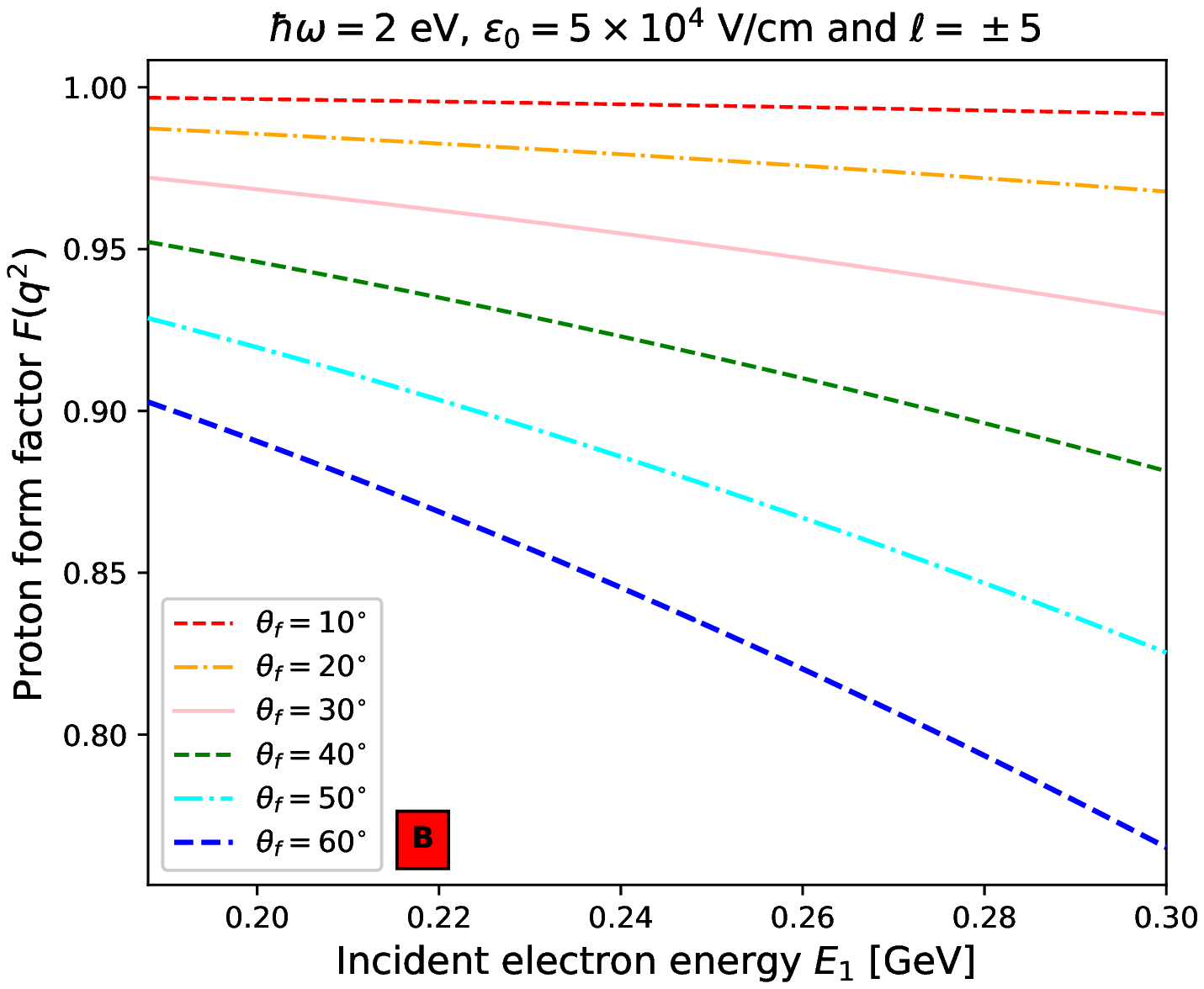}
   \caption{\ref{Figure:5}-A: Form factor $F(q^{2})$ (Eq.(\ref{eq13})) as a function of the laser field strength $\varepsilon_{0}$ for both proton and neutron and for different summations over $\ell$ with $ E_{1}=0.188\,GeV $, $ \theta_{f}=30^{\circ} $ and $\hbar\omega=2\,eV$. 
   \ref{Figure:5}-B: Proton's form factor as a function of $ E_{1} $ for different values of $ \theta_{f} $ by choosing $\ell=\pm5$ and $\hbar\omega=2\,eV$.} \label{Figure:5}
\end{figure}
\begin{table}[H]
\begin{center}
\begin{tabular}{|c|c|c|c|c|}
\toprule
&\multicolumn{2}{|c|}{e-p DCS}&\multicolumn{2}{|c|}{e-n DCS}\\  \cline{1-5}
$E_{1}$ $[\text{MeV}]$ &~~~~~~ F(q$^{2}$) ~~~~~~& $d\bar{\sigma}/d\Omega_{f}$ [GeV$^{-2}$] &~~~~~~ F(q$^{2}$) ~~~~~~& $d\bar{\sigma}/d\Omega_{f}$ [GeV$^{-2}$] \\ \hline
$188$ &~~ 0.972086 ~~& 2.62112$\times10^{-2}$ &~~ 1.00458 ~~& 4.53289$\times10^{-15}$   \\
$200$ &~~ 0.968461 ~~& 2.29493$\times10^{-2}$ &~~ 1.00518 ~~& 5.11073$\times10^{-15}$   \\
$250$ &~~ 0.951060  ~~& 1.40664$\times10^{-2}$ &~~ 1.00804 ~~& 7.86571$\times10^{-15}$   \\
$300$ &~~ 0.930009 ~~& 9.27664$\times10^{-3}$ &~~ 1.01149 ~~& 1.11720$\times10^{-14}$   \\
$350$ &~~ 0.905382 ~~& 6.41533$\times10^{-3}$ &~~ 1.01554 ~~& 1.50180$\times10^{-14}$  \\
 \toprule
\end{tabular}
\end{center}
\caption{Numerical values of form factor and its corresponding DCS as a function of $E_{1}$ for both electron-proton and electron-neutron diffusion. We have chosen the following parameters: $\varepsilon_{0}=5\times10^{4}\,V/cm$, $\hbar\omega=2\,eV$, $\theta_f=30^{\circ}$ and $-5$ $\leq$ $\ell $ $\leq$ $+5$.}\label{table3}
\end{table}
We would like to remind that throughout the electron-nucleon scattering in the absence of a laser field and for large values of the four-momentum transfer $q^{2}$, the electron no longer sees the entire charge of nucleon as in Mott scattering, but rather just a portion of it. In this case, the spatial extension of the nucleon is described by a form factor $F(q^{2})$. When the electron has initial energy $E_{1}=0.188\,GeV$, the electric form factor for the nucleon without spin is $9.72086\times10^{-1}$ for the proton and $1.00458$ for the neutron. 
Figure \ref{Figure:5}-A, in which the pulsed He$:$Ne laser field is applied with a strength $\varepsilon_{0}$ varying from $ 10^{3}\,V/cm $ to $ 10^{8}\,V/cm $, shows that the form factor for both the (e-p) and (e-n) scattering remains constant whatever the number of photons exchanged $\ell$.
However, in Figure \ref{Figure:5}-B ($\varepsilon_{0}=5\times10^{4}\,V/cm $ and $\ell=\pm5$) and for large values of $\theta_{f}$, the form factor decreases considerably when we increase the incident electron's energy in the (e-p) scattering from $0.188\,GeV$ to $0.300\,GeV$. 
This result is interpreted by the fact that scattering with a large angle $\theta_{f}$ corresponds to a large energy transfer.
Therefore, the reduced de-Broglie wavelength $ \bar{\lambda}=\hbar/|\vec{q}|$ that corresponds to the boson exchanged during the scattering process decreases, and this affects the proton form factor. 
In table \ref{table3}, we give some numerical values of the form factor for nucleons scattered by electron impact with a scattering angle $\theta_{f}=30^{\circ}$. The electron is dressed by a laser field with frequency $\hbar\omega=2\,eV$ and strength $\varepsilon_{0}=5\times10^{4}\,V/cm $, and the number of exchanged photons is chosen as $\ell=\pm5$. 
For the proton, and as in figure \ref{Figure:5}-B, we notice that the form factor decreases as well as the differential cross section when the energy of the incident electron increases. On the contrary, the form factor for (e-n) scattering increases as well as the differential cross section by increasing the energy of the incident electron. 
 To conclude, for both (e-p) and (e-n) scattering processes and for a fixed incident electron energy, the laser field modifies the DCS whenever the number of photons exchanged is lower than the cut-off.
\section{Conclusion}\label{Sec3}
In this theoretical study, we firstly presented the electron-nucleon scattering without an external effect, where the nucleon is considered as a spinless particle with a spatial structure described by an electric form factor. Then, we considered this process under the effect of a monochromatic laser field of circular polarization by dressing only the electron in the initial and final states, which are presented as Dirac-Volkov states. We have discussed this scattering process in the cases where the nucleon is considered as a proton or a neutron. We have found that the differential cross section of both (e-p) and (e-n) scattering decreases with the increasing laser field strength, while it increases by increasing the number of photons $ \ell $.
This differential cross section is always lower than its corresponding DCS in the absence of the laser field unless we sum over a number of photons equal or greater than the laser cutoff. Afterward, we concluded that the electric form factor remains constant though we increase the laser field strength from $10^{3}\,V/cm$ to $10^{8}\,V/cm$, while it depends on the energy of the incident electron. Indeed, increasing $E_{1}$ induces a diminution of the form factor in (e-p) diffusion, and it increases that of the (e-p) scattering. 
This study was conducted for nucleons without spin, but it might expand to include nucleons with spin and laser field with linear or elliptic polarization.

\end{document}